\documentclass[preprint,12pt,authoryear]{elsarticle}

\usepackage{amssymb,microtype,paralist}
\usepackage[top=1in, bottom=1in, left=1in, right=1in]{geometry}

\usepackage{amsmath}
\usepackage{graphicx,setspace}
\usepackage{natbib}
\usepackage{url} 
\RequirePackage[OT1]{fontenc}
\RequirePackage{amsthm,amsmath,bm}
\RequirePackage{natbib}
\usepackage{graphicx,subcaption}
\usepackage[pdftex,colorlinks=true]{hyperref}
\hypersetup{citecolor=darkblue,linkcolor=darkblue,urlcolor=darkblue}
\usepackage[dvipsnames,svgnames,x11names]{xcolor}
\definecolor{darkblue}{rgb}{0,0,.6}
\usepackage[draft]{changes}
\definechangesauthor[color=Red]{JCT}
\definechangesauthor[color=DarkGreen]{HH}
\definechangesauthor[color=Brown]{HLS}
\definechangesauthor[color=Purple]{MM}
\usepackage{url}
\usepackage{float}
\usepackage{amsthm}
\usepackage{amsfonts}
\usepackage{amsmath,mathtools,tikz}
\usepackage{comment}
\usepackage{enumerate}
\usepackage{natbib}
\usepackage{animate,dsfont}
\usepackage{multirow}
\usepackage{threeparttable,booktabs}
 \usepackage[figuresright]{rotating}
 \usepackage[pdftex,colorlinks=true]{hyperref}
\definecolor{darkblue}{rgb}{0,0,.6}
\hypersetup{citecolor=darkblue,linkcolor=darkblue,urlcolor=darkblue}

\DeclareMathAlphabet\mathbfcal{OMS}{cmsy}{b}{n} 
\numberwithin{equation}{section}
\theoremstyle{plain}

\newsavebox\CBox

\graphicspath{{plots/}}
\newcommand{\X}{\mathcal{X}}
\DeclareMathOperator*{\argmin}{\arg\!\min}
\usetikzlibrary[decorations.shapes]
\usetikzlibrary[petri]

\journal{Journal of Forecasting}

\begin{document}

\begin{frontmatter}



\title{Forecasting intraday financial time series with sieve bootstrapping and dynamic updating}


\author[MacU]{Han Lin Shang}
\ead{hanlin.shang@mq.edu.au}
\cortext[correspondingauthor]{Corresponding author}

\affiliation[MacU]{organization={Department of Actuarial Studies and Business Analytics, Macquarie University},
          addressline={4 Eastern Road},
	 city =  {Sydney},
	 state = {New South Wales},
	 postcode = {2109},
            country={Australia}}

\author[USYD]{Kaiying Ji}
\affiliation[USYD]{organization={Discipline of Accounting, Governance and Regulation, The University of Sydney},
            addressline={H69 Codrington Building},
             city={Sydney},
             state={New South Wales},
             postcode={2006},
             country={Australia}}

\begin{abstract}
Intraday financial data often take the form of a collection of curves that can be observed sequentially over time, such as intraday stock price curves. These curves can be viewed as a time series of functions observed on equally spaced and dense grids. Due to the curse of dimensionality, high-dimensional data poses challenges from a statistical aspect; however, it also provides opportunities to analyze a rich source of information so that the dynamic changes within short-time intervals can be better understood. We consider a sieve bootstrap method of \cite{PS22} to construct one-day-ahead point and interval forecasts in a model-free way. As we sequentially observe new data, we also implement two dynamic updating methods to update point and interval forecasts for achieving improved accuracy. The forecasting methods are validated through an empirical study of 5-minute cumulative intraday returns of the S\&P/ASX All Ordinaries Index.  
\end{abstract}



\begin{keyword}
Function-on-function linear regression \sep Functional principal component analysis \sep High-frequency financial data \sep Penalized least squares \sep Vector autoregressive model 



\end{keyword}

\end{frontmatter}


\section{Introduction}

High-frequency financial data are observations on financial instruments taken at a finer time scale, such as time-stamped transaction-by-transaction or tick-by-tick data. Advances in computer recording and storage facilitate the presence of such data, which is often used to study market microstructure related to price discovery and market efficiency \citep[see, e.g.,][]{Engle00}. Several pioneer articles on the application of high-frequency financial data sets are provided by \cite{Andersen00}, \cite{CLM97}, \cite{DGM+01}, \cite{Ghysels00}, \cite{GO97}, \cite{GJ01}, \cite{Lyons01}, \cite{Wood00}, \cite{Tsay10}, and \cite{PT21}. This study extends the existing literature by presenting a functional time series method with sieve bootstrapping to obtain well-calibrated interval forecasts. It applies the methodology to the S\&P/ASX All Ordinaries Index (ASX code: XAO).

High-frequency financial data possess unique opportunities absent in low-frequency data. Traditional financial models, such as the celebrated capital asset pricing model \citep{Sharpe64, Lintner65}, consider stock return using low-frequency spot prices, where the price changes between the two spots are largely ignored. If prices are modeled as a univariate time series of discrete observations, the underlying process that generates these observations cannot be fully discovered. Suppose the closing prices are the same for two consecutive days, indicating no movement in the stock market. In contrast, the high-frequency intraday information is particularly interesting in this study because it contains finer trading information and allows us to investigate the movement in an underlying stock. One can study the patterns hidden in the intraday price curves and how the shapes of the intraday price curves change daily.

A high-frequency intraday financial data set is an example of (dense) functional data, represented in a graphical form of curves. Functional time series arise in many fields, including finance \citep[see, e.g.,][]{WJS08, KZ12, KRS17}. Functional time series can be of two types: On one hand, functional time series can arise by separating a continuous time record into natural (subjective) consecutive intervals. On the other hand, functional time series can also arise when observations in a period are considered as finite realizations of a continuous function, such as maturity-specific yield curves \citep[see, e.g.,][]{HSH12}. In either case, the object of interest is a discrete time series of functions with a continuum. The continuum can either be time (e.g., intraday) or other continuous variables (e.g., maturity). In Section~\ref{sec:2}, we describe our data set -- 5-minute cumulative intraday returns of the XAO. This data set belongs to the first type of functional time series.

The advantages of functional time series include: 
\begin{inparaenum}
\item[1)] we can study the temporal correlation of an intraday functional object and learn about how the correlation progress over days; 
\item[2)] we can handle missing values via interpolation or smoothing techniques; 
\item[3)] although not the focus of this paper, we can study not only the level but also the derivatives of functions as a means of reflecting the speed of change \citep[see][]{Shang19, HS22};
\item[4)] by converting a univariate time series in a real-valued space to a time series of functions in function space, we overcome the ``curse of dimensionality" due to natural ordering.
\end{inparaenum}

To model a time series of functions, parametric and nonparametric techniques exist in the statistical literature. In the parametric tools, \cite{Bosq00} proposed a functional autoregressive of order 1 (FAR(1)) and derived one-step-ahead forecasts based on a regularized form of Yule-Walker equations. The FAR(1) model was later extended to a FAR($p$) model by \cite{KR13}, where the order $p$ could be determined using a sequential testing procedure. \cite{KK17} proposed a functional moving average process and introduced an innovations algorithm. By combining autoregressive and moving average, \cite{KKW17} proposed a functional autoregressive and moving average model. While these techniques are designed for analyzing short-memory stationary functional time series, \cite{LRS20, LRS21, LRS22} considered long-range dependent functional time series and proposed a functional autoregressive fractionally integrated moving average model.

In the nonparametric tools, functional principal component analysis has been studied extensively to summarize the characteristics of an infinite-dimensional object by a set of finite-dimensional bases. The functional principal component analysis decomposes a time series of functions into a set of functional principal components and their associated principal component scores. The temporal dependence in a functional time series is inherited by the correlation among principal component scores. We adopt the approach of \cite{ANH15}, who applied a multivariate time series forecasting method, such as a vector autoregressive model, to model and forecast multiple sets of estimated principal component scores jointly. Point forecasts can be obtained based on the historical functions and estimated functional principal components. Since this method uses a time series forecasting method, it is referred to as the ``TS method" in Section~\ref{sec:3.1}.

To quantify forecast uncertainty, a sieve bootstrap procedure of \cite{PS22} is implemented for constructing pointwise prediction intervals and uniform prediction bands. As \cite{Box76} stated ``all models are wrong, but some are useful". Under this principle, point forecasts obtained from a statistical time-series model are a proxy of future expectations and are of little use for purposes such as risk management. The sieve bootstrap in Section~\ref{sec:3.2} takes account of model misspecification error as an additional source of uncertainty, thus improving calibration. The success of this sieve bootstrap is constructing a prediction error distribution that includes the model misspecification error and thus achieves improved calibration.

Forecasting the expected values of stock returns is a questionable exercise. According to \citeauthor{Fama70}'s \citeyearpar{Fama70} weak form of efficient markets hypothesis, forecasting stock returns based on historical data is impossible. However, several studies have shown that forecasting intraday returns are possible. In a cross-section of stocks, \cite{HKS10} demonstrated a pattern of return continuation at half-hour intervals that were multiples of a trading day, and this pattern can last for at least 40 trading days. \cite{GHL+17} demonstrated that first half-hour returns on the market predict the last half-hour returns. This study extends the findings of \cite{GHL+17} and demonstrates that dynamic updating methods can be applied throughout the day with various degrees of accuracy and can be generalized to other capital markets, such as Australia.

When partially observed intraday data in the most recent curve becomes available, we should incorporate this new information to improve forecast accuracy. We present two dynamic updating methods, namely penalized least squares and function-on-function linear regression, for updating point and interval forecasts in Section~\ref{sec:4}. The dynamic updating methods can incorporate any intraday price jumps and obtain revised point and interval forecasts. By inputting sieve bootstrap forecasts into the dynamic updating methods, we may obtain improved forecast accuracy, which is one of the primary contributions of the paper. Using the forecast error measures in Section~\ref{sec:5}, via the cumulative intraday returns (CIDRs) of the XAO, we evaluate and compare point and interval forecast accuracies in Section~\ref{sec:6}. Conclusions are given in Section~\ref{sec:7}, along with some ideas on how the methodology can be further extended.

\section{Cumulative intraday returns of S\&P/ASX All Ordinaries (XAO)}\label{sec:2}

The All Ordinaries is designed to measure the 500 largest companies in the Australian equities market, drawn from eligible companies listed on the Australian Securities Exchange. According to the Union Bank of Switzerland, the Australian market is highly concentrated, with large financial, resources, and technology companies. As the first major financial market to open each day, the Australian Securities Exchange is a world leader in raising capital and provides active managers with opportunities to beat the market. We focus on XAO to add some fresh evidence to the existing forecasting literature, which largely overlooks such a major capital market. The 5-minute intraday close prices of XAO from January 4 to December 23, 2021, were downloaded from Refinitiv (\url{https://select.datascope.refinitiv.com/DataScope/}). We consider one year of data to avoid any structural change. Following the work of \cite{HL06}, we also consider a 5-minute sampling frequency which is a popular choice when avoiding the adverse effects of microstructure noise. 

Let $P_t(u_i)$, $t=1,2,\dots,n$, $i=2, 3, \dots,\tau$ and $\tau=75$ be the 5-minute close price of the XAO at intraday time $u_i$ between 10:00 and 16:10 Sydney time on day $t$. The close time of 16:10 is to allow all trades that take place on the close transaction at the same price regardless of what price an investor bids or offers. Any overnight trading will be reflected at the beginning close price the next day. For a stationary transformation, we compute CIDRs as discussed in \cite{KRS17}, defined as
\begin{equation}
\X_t(u_i) = 100\times [\ln P_t(u_i) - \ln P_t(u_1)], \quad i=2, 3, \dots,\tau,\label{eq:CIDR}
\end{equation}
where $\ln(\cdot)$ represents the natural logarithm. There are other versions of intraday return curves \citep[see][]{KRS17, RWZ20}. We choose the CIDR curve to eliminate the possible jump at the first close price. The CIDR curves preserve intraday movements, enhancing our understanding of the evolutions of intraday returns for XAO. From~\eqref{eq:CIDR}, we take the inverse transformation to obtain the 5-minute intraday price
\[
P_t(u_i) = \exp^{\frac{\X_t(u_i)}{100}}\times P_t(u_1).
\]

\begin{figure}[!htb]
\centering
\subfloat[Univariate time series plot]
{\includegraphics[width=8cm]{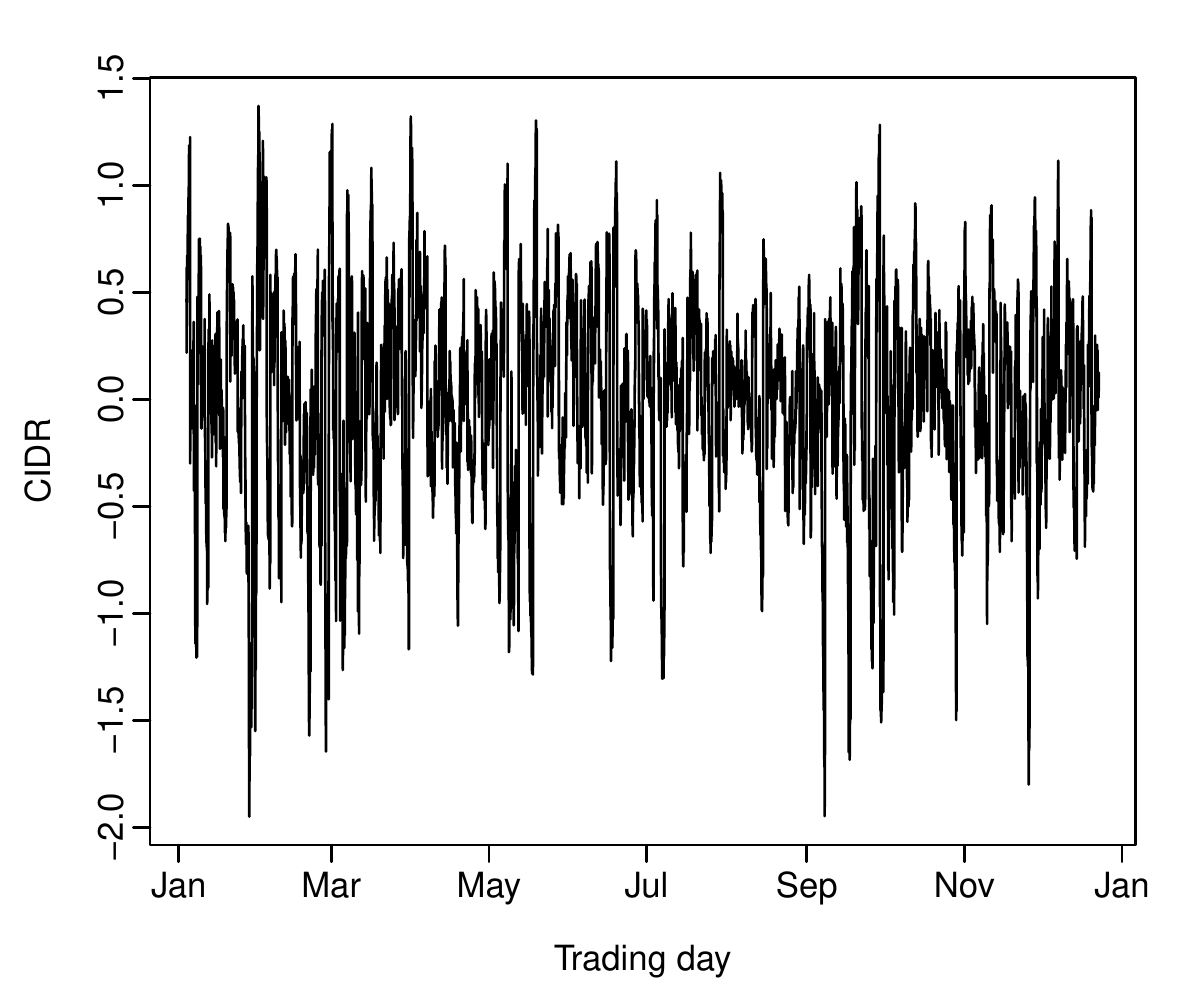}\label{fig:1a}}
\quad
\subfloat[Functional time series plot]
{\includegraphics[width=8cm]{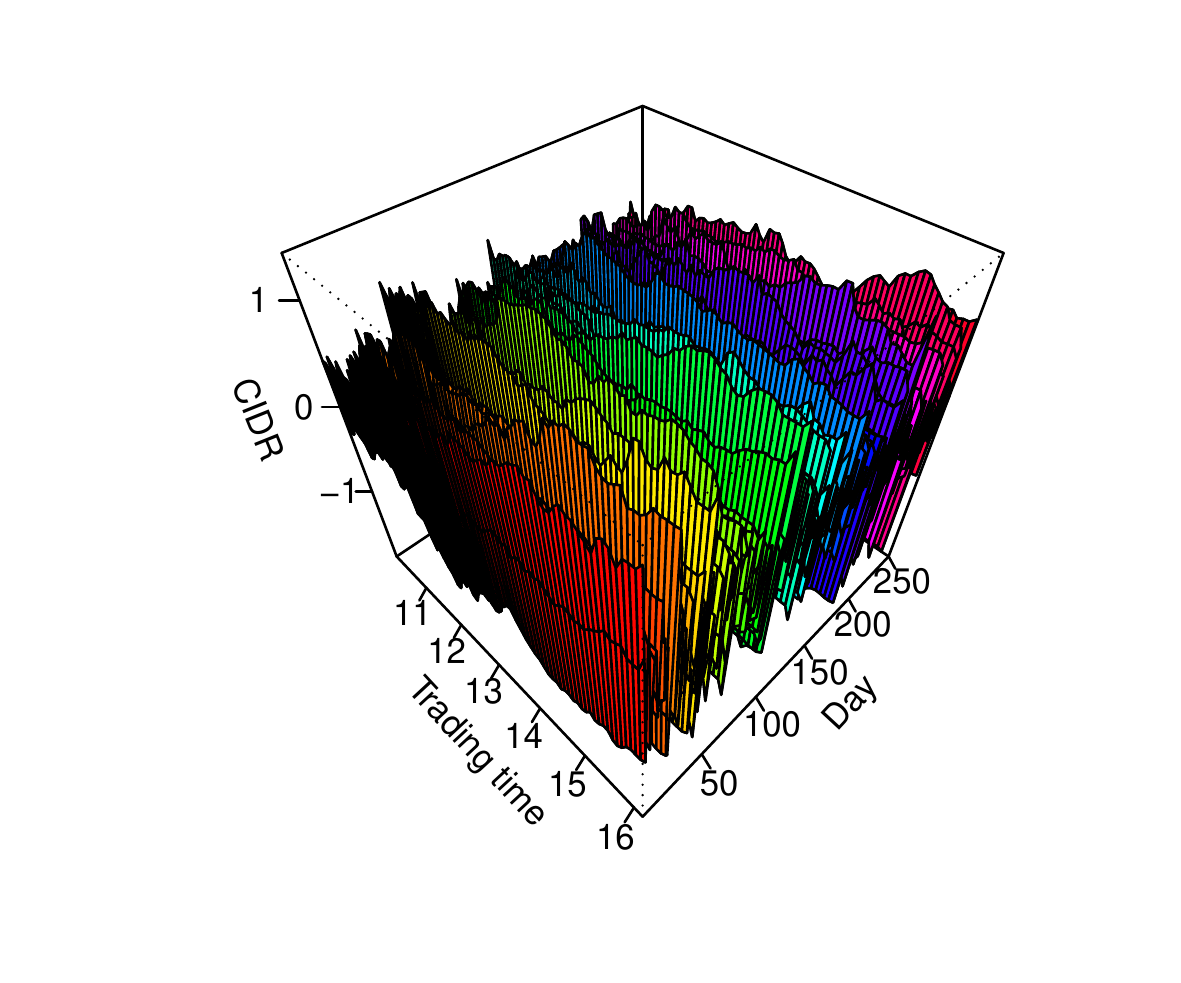}\label{fig:1b}}
\caption{Graphical displays of CIDRs of the XAO from January 4 to December 23, 2021.}\label{fig:1}
\end{figure}

Since $\tau>\lfloor \sqrt{n}\rfloor$, we have a dense functional data setting \citep{ZW16}. Via a linear interpolation, continuous curves can be constructed from these 74 CIDRs and we form a time series of functions $\X_t(u)$. In the XAO data set, there are 74 five-minute CIDRs, representing around 6 hours of trading at the Australian Securities Exchange on a trading day. A univariate time series of $18,500=74\times 250$ discrete returns was converted into $n=250$ days of CIDRs. Figure~\ref{fig:1a} presents a univariate time series display of CIDRs, whereas Figure~\ref{fig:1b} displays the same data as a time series of functions. From Figure~\ref{fig:1a}, the data series appears to have a constant mean, persistence and no apparent volatility clustering. Through a functional KPSS test of \cite{HKR14}, the series is trend stationary with a $p$-value of 0.737, implying no presence of a unit root. 

\section{Functional time series forecasting method}\label{sec:3}

\subsection{Functional principal component regression}\label{sec:3.1}

Let $(\X_t, t\in \mathbb{Z})$ be an arbitrary stationary functional time series. Each function is a square-integrable function in Hilbert space $\mathcal{H}$ satisfying $\|\X_t\|^2 = \int_{\mathcal{I}}\X_t^2(u)du<\infty$ where $\mathcal{I}\in [u_2, u_{\tau}]$ denotes a function support. While $u_2$ symbolizes the first value of CIDRs recorded at 10:05, and $u_{\tau}$ denotes the last value of CIDRs recorded at 16:10 on a trading day. For a time series of functions $[\X_1(u), \X_2(u),\dots, \X_n(u)]$, the mean function can be estimated by $\overline{\X}(u) = \frac{1}{n}\sum^n_{t=1}\X_t(u)$. The covariance function can be estimated by
\[
\text{cov}[\X(u), \X(v)]=\text{E}\left\{[\X(u) - \overline{\X}(u)][\X(v) - \overline{\X}(v)]\right\}.
\]
Via Mercer's lemma, the covariance function can be approximated by orthonormal eigenfunctions,
\[
\text{cov}[\X(u), \X(v)] = \sum^{\infty}_{k=1}\widehat{\lambda}_k \widehat{\phi}_k(u)\widehat{\phi}_k(v), \qquad u, v\in \mathcal{I}, 
\]
where $\widehat{\phi}_k(u)$ denotes the $k$\textsuperscript{th} orthonormal functional principal component, $\widehat{\lambda}_k$ denotes the $k$\textsuperscript{th} eigenvalue, and $\widehat{\lambda}_1\geq \widehat{\lambda}_2\dots> 0$.

Orthonormal eigenfunctions can well approximate realizations of a stochastic process. Via the Karhunen-Lo\`{e}ve expansion, the functional realization $\X_t(u)$ can be expressed as
\begin{equation}
\X_t(u) = \overline{\X}(u)+\sum^{\infty}_{k=1}\widehat{\beta}_{t,k}\widehat{\phi}_k(u),\label{eq:2}
\end{equation}
where $\widehat{\beta}_{t,k}$ is obtained by projecting $[\X_t(u) - \overline{\X}(u)]$ onto the direction of the $k$\textsuperscript{th} estimated functional principal component $\widehat{\phi}_k(u)$, i.e., $\widehat{\beta}_{t,k}=\langle \X_t(u) - \overline{\X}(u), \widehat{\phi}_k(u)\rangle$.

To achieve dimensionality reduction, we truncate~\eqref{eq:2} to the first $K$ terms. Because eigenvalues are ordered in a non-increasing pattern, the first few terms often capture main mode of information in the data. Without losing much information, $\X_t(u)$ can be adequately captured by the $K$-dimensional vector $(\bm{\widehat{\beta}}_1, \bm{\widehat{\beta}}_2, \dots,\bm{\widehat{\beta}}_K)$, where $\bm{\widehat{\beta}}_1 = (\widehat{\beta}_{1,1}, \widehat{\beta}_{1,2}, \dots,\widehat{\beta}_{1,n})$. We obtain
\begin{equation}
\X_t(u) = \overline{\X}(u) + \sum^K_{k=1}\widehat{\beta}_{t,k}\widehat{\phi}_k(u) + e_t(u),\label{eq:FPCA}
\end{equation}
where $e_t(u)$ represents the error terms containing estimated functional principal components and their associated scores excluded from the first $K$ terms. The number of retained components, $K$, was chosen as 
\begin{equation}
K=\argmin_{1\leq k\leq k_{\max}}\left\{\frac{\widehat{\lambda}_{k+1}}{\widehat{\lambda}_k}\times \mathds{1}(\frac{\widehat{\lambda}_k}{\widehat{\lambda}_1}\geq \upsilon) + \mathds{1}(\frac{\widehat{\lambda}_k}{\widehat{\lambda}_1}< \upsilon)\right\}, \label{eq:eigenvalue_ratio}
\end{equation}
where $\upsilon = 1/\ln[\max(\widehat{\lambda}_1, n)]$ is a pre-specified positive number, $k_{\max}=\#\{k|\widehat{\lambda}_k\geq \sum^n_{k=1}\widehat{\lambda}_k/n\}$, and $\mathds{1}\{\cdot\}$ represents the binary indicator function. 


The sets of estimated principal component scores, $(\bm{\widehat{\beta}}_1, \bm{\widehat{\beta}}_2, \dots,\bm{\widehat{\beta}}_K)$, can be modeled and forecasted via a multivariate time series forecasting method, such as vector autoregressive of order $p$, denoted by VAR($p$),
\begin{equation}
\bm{\widehat{\beta}}_t = \sum^p_{\xi=1}\bm{\widehat{A}}_{\xi,p}\bm{\widehat{\beta}}_{t-\xi}+\bm{\widehat{\epsilon}}_t,\label{eq:VAR_forward}
\end{equation}
where $t=p+1,p+2,\dots,n$, $\bm{\widehat{A}}_{\xi, p}$ is an estimated $(K\times K)$ coefficient matrix of the forward score series, and $(\bm{\widehat{\epsilon}}_{p+1}, \bm{\widehat{\epsilon}}_{p+2}, \dots,\bm{\widehat{\epsilon}}_n)$ are the residuals obtained by fitting the VAR($p$) model to the $K$-dimensional multivariate time series of estimated scores. The order $p$ of a fitted VAR model can be chosen using a corrected Akaike information criterion (AIC$_{\text{c}}$) of \cite{HT93} by minimizing
\begin{equation}
\text{AIC}_{\text{c}}(p) = n\ln|\widehat{\bm{\Sigma}}_{\widehat{\bm{\epsilon}},p}|+\frac{n(nK+p K^2)}{n-K(p+1)-1},\label{eq:AICC}
\end{equation}
over a set of possible values $p=\{1,2,\dots,10\}$, and $\widehat{\bm{\Sigma}}_{\widehat{\bm{\epsilon}},p}=\frac{1}{n-p}\sum^n_{t=p+1}\widehat{\bm{\epsilon}}_{t}\widehat{\bm{\epsilon}}^{\top}_{t}$ and $^{\top}$ represents matrix transpose.

Conditional on observed time series of functions $\bm{\X}(u)=[\X_1(u), \X_2(u), \dots,\X_n(u)]$, estimated mean function $\overline{\X}(u)$ and estimated functional principal components $\bm{\Phi}(u) = [\widehat{\phi}_1(u), \widehat{\phi}_2(u), \dots,\widehat{\phi}_K(u)]$, the one-step-ahead point forecast can be obtained as
\begin{align*}
\widehat{\X}_{n+1|n}(u) &= \text{E}[\X_{n+1}(u)|\bm{\X}(u), \overline{\X}(u), \bm{\Phi}(u)] \\
&=\overline{\X}(u) + \sum^K_{k=1}\widehat{\beta}_{n+1|n,k}\widehat{\phi}_k(u),
\end{align*}
where $\widehat{\phi}_k(u)$ denotes the $k$\textsuperscript{th} estimated functional principal component, and $\widehat{\beta}_{n+1|n,k}$ represents the one-step-ahead point forecasts of $\beta_{n+1,k}$ using the VAR($p$) model. When $K=1$, we only have a univariate time series of principal component scores, which can be modeled and forecasted via an autoregressive (AR) model of order $p$.

\subsection{Sieve bootstrap}\label{sec:3.2}

For every $t\in \mathbb{Z}$, the zero-mean random element $\X_t$ is generated as
\[
\X_t = f(\X_{t-1}, \X_{t-2},\dots) + \varepsilon_t,
\]
where $f:\mathcal{H}^{\infty}\rightarrow \mathcal{H}$ is some appropriate operator and $\{\varepsilon_t\}$ is a zero-mean independent and identically distributed (i.i.d.) innovation process in $\mathcal{H}$ with $\text{E}\|\varepsilon_t\|^2<\infty$ and $\|\cdot\|$ the norm of $\mathcal{H}$.

Based on the last $\ell$ observed functions, $\X_{n,\ell}=(\X_n, \X_{n-1}, \dots, \X_{n-\ell+1})$, for $\ell<n$, a predictor
\[
\widehat{\X}_{n+1} = \widehat{g}(\X_n, \X_{n-1}, \dots, \X_{n-\ell+1}),
\]
of $\X_{n+1}$ is used, where $\widehat{g}: \mathcal{H}^{\ell}\rightarrow \mathcal{H}$ some estimated operator.

The prediction error $\mathcal{E}_{n+1}=\X_{n+1}-\widehat{\X}_{n+1}$ given $\X_{n,\ell}$ can be expressed as
\begin{align*}
\mathcal{E}_{n+1}=&\ \X_{n+1} - \widehat{\X}_{n+1} \\
=&\ \vartheta_{n+1} + [f(\X_{n}, \X_{n-1}, \dots) - g(\X_{n}, \X_{n-1},\dots, \X_{n+1-\ell})] + \\
&[g(\X_{n}, \X_{n-1}, \dots, \X_{n+1-\ell}) - \widehat{g}(\widehat{\X}_{n}, \widehat{\X}_{n-1}, \dots, \widehat{\X}_{n+1-\ell})] \\
=& \ \mathcal{E}_{I, n+1} + \mathcal{E}_{M, n+1} + \mathcal{E}_{E, n+1},
\end{align*}
where $\mathcal{E}_{I,n+1}$ is the error attributable to the i.i.d. innovation, $\mathcal{E}_{M,n+1}$ is the model misspecification error, and $\mathcal{E}_{E,n+1}$ is the error attributable to estimation of the unknown operator $g$ and of the random elements $(\X_{n}, \dots, \X_{n+1-\ell})$ used for one-step-ahead prediction. Both $\mathcal{E}_{I,n+1}$ and $\mathcal{E}_{E,n+1}$ go to zero asymptotically, but the model misspecification error $\mathcal{E}_{M,n+1}$ may not vanish if the model used for prediction differs from the one generating the data.

We implement a sieve bootstrap method of \cite{PS22} to take into account all three aforementioned sources affecting the prediction error and construct a prediction band for $\X_{n+1}$ associated with the predictor $\widehat{\X}_{n+1}$ in a model-free way. Given $\X_{n,\ell}$ of the functional time series observed and for any $\alpha\in (0,1)$, we construct a prediction band denoted by $\{[\widehat{\X}_{n+1}(u) - L_{n}(u), \widehat{\X}_{n+1}(u) + U_{n}(u)]\}$ such that
\[
\lim_{n\rightarrow \infty}\text{Pr}\left(\widehat{\X}_{n+1}(u) - L_{n}(u)\leq \X_{n+1}(u)\leq \widehat{\X}_{n+1}(u) + U_{n}(u), \forall u\in \mathcal{I}\big|\X_{n,\ell}\right)=1-\alpha,
\]
where $\alpha$ is a level of significance.

The sieve bootstrap uses the VAR($p$) process in~\eqref{eq:VAR_forward} to generate forward score forecasts
\begin{equation}
\bm{\beta}^*_{n+1}=\sum^p_{\xi=1}\widehat{\bm{A}}_{\xi,p}\bm{\beta}^*_{n+1-\xi}+\bm{\epsilon}^*_{n+1},\label{eq:score_forward}
\end{equation}
where $\bm{\beta}^*_{n+1-\xi}=\widehat{\bm{\beta}}_{n+1-\xi}$ for $n+1-\xi\leq n$ and $\bm{\epsilon}^*_{n+1}$ is i.i.d. resampled from the set of centered residuals $\{\widehat{\bm{\epsilon}}_t - \overline{\bm{\epsilon}}, t=p+1,p+2,\dots,n\}$, and $\overline{\bm{\epsilon}}=\frac{1}{n-p}\sum^n_{t=p+1}\widehat{\bm{\epsilon}}_t$. From~\eqref{eq:FPCA}, we compute
\[
\X_{n+1}^*(u)=\overline{\X}(u)+\sum^K_{k=1}\beta^*_{n+1,k}\widehat{\phi}_k(u)+e^*_{n+1}(u),
\]
where $e^*_{n+1}(u)$ are i.i.d. resampled from the set $\{e_{t}(u)-\overline{e}(u), t=1,2,\dots,n\}$ and $\overline{e}(u)=\frac{1}{n}\sum^n_{t=1}e_{t}(u)$ and $e_t(u) = \X_t(u) - \overline{\X}(u) - \sum^K_{k=1}\widehat{\beta}_{t,k}\widehat{\phi}_k(u)$.

Because of stationarity, the VAR($p$) process can also go backward in time to generate bootstrap samples of the estimated scores; that is,
\begin{equation*}
\bm{\widehat{\beta}}_t = \sum^p_{\xi=1}\bm{\widehat{B}}_{\xi,p}\bm{\widehat{\beta}}_{t+\xi}+\bm{\eta}_t,\qquad t=1,2,\dots,n-p,
\end{equation*}
where $\bm{\widehat{B}}_{\xi,p}$ denotes an estimated $(K\times K)$ coefficient matrix for the backward score series, and $\bm{\eta}_t$ is the error term, related to $\bm{\epsilon}_t$ in~\eqref{eq:VAR_forward}. The bootstrap samples $\bm{\eta}^*_t$ can be obtained by
\[
\bm{\eta}_t^* = \bm{B}_p(L^{-1})\bm{A}_p^{-1}(L)\bm{\epsilon}_t^*,
\]
where $\bm{A}_p(z) = \bm{I}_{K} - \sum^p_{\xi=1}\bm{A}_{\xi,p}z^{\xi}$, $\bm{B}_p(z) = \bm{I}_{K} - \sum^p_{\xi=1}\bm{B}_{\xi,p}z^{\xi}$, $\bm{I}_{K}$ denotes a $(K\times K)$ diagonal matrix,  $z\in \mathbb{C}$, $\mathbb{C}$ denotes a complex number, and $\bm{\epsilon}_t^*$ denotes the bootstrapped forward scores in~\eqref{eq:score_forward}.

The bootstrap samples for the backward scores can be obtained via
\[
\bm{\beta}^*_t = \sum^p_{\xi=1}\bm{\widehat{B}}_{\xi,p}\bm{\beta}_{t+\xi}^*+\bm{\eta}_t^*.
\]
From~\eqref{eq:FPCA}, a pseudo-functional time series $\{\X_1^*(u),\X_2^*(u),\dots,\X_n^*(u)\}$ can be obtained via
\begin{equation}
\X_t^*(u) = \overline{\X}(u) + \sum^K_{k=1}\beta_{t,k}^*\widehat{\phi}_{k}(u)+e_{t}^*(u),\label{eq:FPCA_boot}
\end{equation}
where $e_t^*(u)$ are i.i.d. pseudo-element resampled from $\{e_{t}(u)-\overline{e}(u), t=1,2,\dots,n\}$.

For each pseudo functional time series, one can apply any functional time series forecasting method, such as the FAR(1) model of \cite{Bosq00}, to obtain the prediction. The FAR(1) model is given by
\[
\widehat{\X}_{n+1}(u) = \overline{\X}(u) + \gamma[\X_{n}(u) - \overline{\X}(u)],
\]
where $\gamma[\cdot]$ is a bounded linear operator capturing the first-lag autocorrelation. The operator $\gamma$ can be estimated from empirical variance and autocovariance operators
\begin{align*}
\widehat{\gamma}&=\frac{\widehat{\Gamma}(1)}{\widehat{\Gamma}(0)} \\
\widehat{\Gamma}(0)&= \frac{1}{n}\sum^n_{t=1}[\X_t(u)-\overline{\X}(u)]\otimes  [\X_t(u)-\overline{\X}(u)] \\
\widehat{\Gamma}(1)&= \frac{1}{n}\sum^{n-1}_{t=1}[\X_t(u) - \overline{\X}(u)]\otimes [\X_{t+1}(u) - \overline{\X}(u)].
\end{align*}

The distribution of prediction error $\mathcal{E}_{n+1}^*(u) = \X_{n+1}^*(u) - \widehat{\X}_{n+1}^*(u)$ is a good proxy for the distribution of $\mathcal{E}_{n+1}(u) = \X_{n+1}(u) - \widehat{\X}_{n+1}(u)$ given $[\X_{n-\ell+1}(u), \dots,\X_n(u)]$, where $\widehat{\X}_{n+1}(u)$ is the one-step-ahead point forecast obtained by the same functional time series forecasting method, such as the FAR(1) model, applied to the original functional time series.

From $\mathcal{E}_{n+1}^*(u)$, we compute its standard deviation, denoted by $\sigma^*_{n+1}(u)$. For $u\in \mathcal{I}$, we compute a normalized statistic
\[
V^*_{n+1}(u) = \frac{\X_{n+1}^*(u) - \widehat{\X}_{n+1}^*(u)}{\sigma_{n+1}^*(u)}.
\]
As shown in \cite{PS22}, it holds true
\[
\sup_{u\in \mathcal{I}}\left|[\sigma_{n+1}^{*}(u)]^2 - \sigma_{n+1}^2(u)\right|\rightarrow 0, \qquad \text{in probability}.
\]
Thus, $V^{*}_{n+1}(u)$ is a good proxy for the distribution of $V_{n+1}(u) = [\X_{n+1}(u) - \widehat{\X}_{n+1}(u)]/\sigma_{n+1}(u)$.

The $(1-\alpha)$ pointwise prediction interval for $\X_{n+1}(u_i)$ can be constructed as $\alpha/2$ and $(1-\alpha/2)$ quantiles of $[\widehat{\X}_{n+1}(u_i) + \mathcal{E}_{n+1}^*(u_i)]$ for $i=2,\dots,\tau$. Let $M^* = \sup_{u\in \mathcal{I}}|V_{n+1}^*(u)|$, and denote by $Q_{1-\alpha}^*$ the $(1-\alpha)$ quantile of the distribution of $M^*$. The $(1-\alpha)$ uniform prediction band for $\X_{n+1}(u)$ associated with the predictor $\widehat{\X}_{n+1}(u)$, is then given by
\[
\Big[\widehat{\X}_{n+1}(u)-Q_{1-\alpha}^*\sigma_{n+1}^*(u), \quad \widehat{\X}_{n+1}(u)+Q_{1-\alpha}^*\sigma_{n+1}^*(u)\Big].
\]

\section{Updating point and interval forecasts}\label{sec:4}

When a functional time series is formed as segments of a univariate time series, the most recent curve is observed sequentially and may not represent a complete curve. Denote the first $m$ time periods of $\X_{n+1}(u)$ by $\X_{n+1}(u_e) = [\X_{n+1}(u_2), \X_{n+1}(u_3), \dots,\X_{n+1}(u_m)]^{\top}$, we are particularly interested in forecasting the data in the remainder of day $n+1$, denoted by $\X_{n+1}(u_l)$, where $u_l\in (u_m, u_{\tau}]$. In Figure~\ref{fig:2}, we present a conceptual diagram for dynamic updating.

\tikzset{decorate with/.style={fill=cyan!20,draw=cyan}}
\begin{figure}[!htb]
\begin{center}
\scalebox{.7}{
\begin{tikzpicture}
\draw (0,0) node[anchor=north east]{$u_{\tau}$} rectangle (12,8);
\draw (0, 8) node[anchor=north east]{$u_2$} -- (0, 0);
\draw (0,4.1) node[anchor=north east]{  \begin{rotate}{90} \hspace {-.5in} Dimensionality \end{rotate} \hspace{.05in} $u_{m}$} -- (13, 4.1);
\draw (11.86, 8) node[anchor=south west]{$n+1$} -- (13, 8);
\draw (13, 8) -- (13, 4.1);
\draw (8.5,0) node[anchor=north east]{Number of curves} -- (12,0);
\draw(0.4,8) node[anchor=south west]{1} -- (1.1,8);
\draw(1.4,8) node[anchor=south west]{2} -- (2.1,8);
\draw(11.1,8) node[anchor=south west]{$n$} -- (12.1,8);
\end{tikzpicture}}
\end{center}
\caption{Conceptual diagram of dynamic updating.}\label{fig:2}
\end{figure}
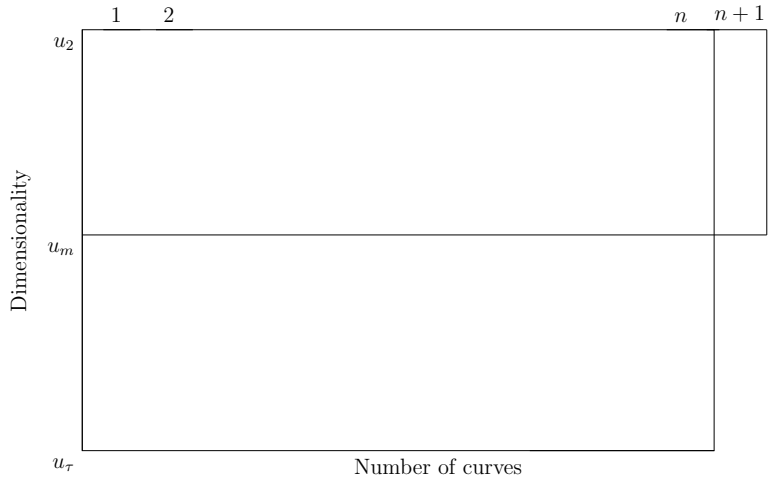

The forecasting method in Section~\ref{sec:3.1} does not use the most recent data. To improve forecast accuracy, it is desirable to dynamically update the point forecasts for the remaining period of the most recent curve $n+1$ by incorporating the newly arrived information \citep[see, e.g.,][]{Chiou12, LXC+18, YY22}. We consider two dynamic updating methods in \cite{Shang17} and \cite{SYK19} that work well with the sieve bootstrap, which is one of the contributions of the present paper. While \cite{Shang17} considered dynamic updating in the S\&P~500 return, \cite{SYK19} studied it in a VIX implied by S\&P 500 index option prices.

\subsection{Penalized least squares (PLS)}\label{sec:4.1}

Let $\X_{n+1}^c(u_e) = \X_{n+1}(u_e)-\overline{\X}(u_e)$. This dynamic updating method shrinks regression coefficient estimates towards $\widehat{\bm{\beta}}^{\text{TS}}_{n+1}$ in~\eqref{eq:VAR_forward}. The PLS regression coefficient estimates minimize a penalized residual sum of squares:
\begin{equation}
\argmin_{\bm{\beta}_{n+1}} \left\{[\X_{n+1}^c(u_e) - \bm{\mathcal{F}}_e\bm{\beta}_{n+1}]^{\top}[\X_{n+1}^c(u_e) - \bm{\mathcal{F}}_e\bm{\beta}_{n+1}]+\lambda(\bm{\beta}_{n+1}-\widehat{\bm{\beta}}_{n+1|n}^{\text{TS}})^{\top}(\bm{\beta}_{n+1}-\widehat{\bm{\beta}}_{n+1|n}^{\text{TS}})\right\}.\label{eq:PLS}
\end{equation}
where $\lambda\in (0, \infty)$ is a shrinkage parameter, and $\bm{\mathcal{F}}_e$ be the $m\times K$ matrix, whose $(i,k)$\textsuperscript{th} entry is $\widehat{\phi}_k(u_i)$ for $2\leq i\leq m$ and $1\leq k\leq K$. The first term in~\eqref{eq:PLS} measures the in-sample goodness-of-fit, while the second term penalizes the departure of the regression coefficient estimates from $\widehat{\bm{\beta}}_{n+1|n}^{\text{TS}}$. By taking the first derivative with respect to $\bm{\beta}_{n+1}$, we obtain
\begin{equation}
\widehat{\bm{\beta}}_{n+1}^{\text{PLS}}=(\bm{\mathcal{F}}_e^{\top}\bm{\mathcal{F}}_e+\lambda\bm{I}_K)^{-1}\left[\bm{\mathcal{F}}_e^{\top}\X^c_{n+1}(u_e)+\lambda\widehat{\bm{\beta}}_{n+1|n}^{\text{TS}}\right].\label{eq:PLS_2}
\end{equation}
When the shrinkage parameter $\lambda\rightarrow 0$, $\widehat{\bm{\beta}}_{n+1}^{\text{PLS}}$ approach to $\widehat{\bm{\beta}}_{n+1}^{\text{OLS}} = (\bm{\mathcal{F}}_e^{\top}\bm{\mathcal{F}}_e)^{-1}\bm{\mathcal{F}}_e^{\top}\X^c_{n+1}(u_e)$ given $(\bm{\mathcal{F}}_e^{\top}\bm{\mathcal{F}}_e)^{-1}$ exists; when $\lambda\rightarrow \infty$, $\widehat{\bm{\beta}}_{n+1}^{\text{PLS}}$ approach to $\widehat{\bm{\beta}}_{n+1|n}^{\text{TS}}$; when $0<\lambda<\infty$, $\widehat{\bm{\beta}}_{n+1}^{\text{PLS}}$ is a weighted average between $\widehat{\bm{\beta}}_{n+1|n}^{\text{TS}}$ and $\widehat{\bm{\beta}}_{n+1}^{\text{OLS}}$. With an optimal selection of $\lambda$, the PLS forecasts of $\X_{n+1}(u_l)$ at discretized time points can be given by
\[
\widehat{\X}_{n+1}^{\text{PLS}}(u_l) = \overline{\X}(u_l)+\sum^K_{k=1}\widehat{\beta}_{n+1,k}^{\text{PLS}}\widehat{\phi}_k(u_l).
\]

The finite-sample performance of the PLS method crucially depends on the optimal selection of $\lambda$. The optimal values of $\lambda$ need to vary intraday to achieve minimal forecast errors. To select $\lambda$, we consider a holdout forecast evaluation. We divide a time series of functions into a training sample consisting of curves one to 200 and a testing sample consisting of curves 201 to 250. The training sample size covers 80\% of the total amount of the data, while the remaining 20\% is the testing sample. The training data cover about ten months of trading, from 4 January 2021 to 14 October 2021. Within the training sample, we further divide the data into a training set consisting of curves one to 150 and a validation set consisting of curves 151 to 200. Within the validation set, the optimal values of $\lambda$ for different updating periods were determined by minimizing the averaged forecast error criteria in Section~\ref{sec:5}. We implicitly assume that the $\lambda$ values selected from the validation set do not differ much from the $\lambda$ values in the testing samples.

When we have partially observed data, we can also dynamically update prediction intervals through the sieve bootstrap in Section~\ref{sec:3.2}. We bootstrap $B$ samples of the TS forecast regression coefficient, denoted by $\widehat{\bm{\beta}}_{n+1|n}^{*,\text{TS}}=\left(\widehat{\beta}_{n+1|n,1}^{*,\text{TS}}, \widehat{\beta}_{n+1|n,2}^{*,\text{TS}}, \dots,\widehat{\beta}_{n+1|n,K}^{*,\text{TS}}\right)^{\top}$. From~\eqref{eq:PLS_2}, these bootstrap TS regression coefficient estimates lead to $\widehat{\bm{\beta}}_{n+1}^{*,\text{PLS}}$. From $\widehat{\bm{\beta}}_{n+1}^{*,\text{PLS}}$, we obtain $B$ replications of 
\[
\widehat{\X}_{n+1}^{*,\text{PLS}}(u_l) = \overline{\X}(u_l)+\sum^K_{k=1}\beta_{n+1,k}^{*,\text{PLS}}\widehat{\phi}_k(u_l)+e^*_{n+1}(u_l),
\]
where $e_{n+1}^*(u_l)$ denotes the bootstrapped residuals corresponding to the updating period, see~\eqref{eq:FPCA_boot}. The $(1-\alpha)$ prediction intervals for the updated forecasts are defined as $\alpha/2$ and $(1-\alpha/2)$ empirical quantiles of $\left\{\widehat{\X}_{n+1}^{1,\text{PLS}}(u_l), \widehat{\X}_{n+1}^{2,\text{PLS}}(u_l), \dots, \widehat{\X}_{n+1}^{B,\text{PLS}}(u_l)\right\}$, where $B=400$ denotes the number of bootstrap samples.
 
\subsection{Function-on-function linear regression}\label{sec:4.2}

By discretizing a function into a set of grids, the PLS method updates forecasts for the updating period. Dynamic updating can also be achieved via a function-on-function linear regression of \citet[][Chapter 16]{RS05}, given by
\[
\X_{n+1}^l(u) = \overline{\X}^l(u)+\int_{v}[\X_{n+1}^e(v) - \overline{\X}^e(v)]\beta(u,v)dv + \zeta_{n+1}^l(u),
\]
where $v\in[u_2, u_m]$ and $u\in (u_m, u_{\tau}]$ represent two function support ranges for the partially observed and updating data periods; $\overline{\X}^e(v)$ and $\overline{\X}^l(u)$ represent two mean functions for the partially observed and updating data periods; $\X_{n+1}^e(v)$ and $\X_{n+1}^l(u)$ denote the functional predictor and response, respectively; and $\beta(u,v)$ denotes a bivariate regression coefficient function.

The time series of functions is divided into two parts: one block contains all data up to the most recent intraday period, and the other contains all data in the updating period. Via the functional principal component analysis, we obtain

\begin{minipage}{0.5\textwidth}
\begin{align*}
\X_t^e(v)&=\overline{\X}^e(v)+\sum^{\infty}_{r=1}\widehat{\theta}_{t,r}\widehat{\phi}_r^e(v) \\
&=\overline{\X}^e(v)+\sum^{R}_{r=1}\widehat{\theta}_{t,r}\widehat{\phi}_r^e(v)+\kappa_t^e(v) \\
\end{align*}
\end{minipage}
\begin{minipage}{0.5\textwidth}
\begin{align*}
\X_t^l(u)&=\overline{\X}^l(u)+\sum^{\infty}_{s=1}\widehat{\vartheta}_{t,s}\widehat{\phi}_s^l(u) \\
&=\overline{\X}^l(u)+\sum^{S}_{s=1}\widehat{\vartheta}_{t,s}\widehat{\phi}_s^l(u)+\delta_t^l(u) \\
\end{align*}
\end{minipage}
where $\widehat{\phi}_r^e(v)$ and $\widehat{\phi}_s^l(u)$ denote the $r$\textsuperscript{th} and $s$\textsuperscript{th} estimated functional principal components associated with the partially observed and updating data periods; $\widehat{\theta}_{t,r}$ and $\widehat{\vartheta}_{t,s}$ are the principal component scores of $\X_t^e(v)$ and $\X_t^l(u)$; $R$ and $S$ are the retained numbers of components selected by~\eqref{eq:eigenvalue_ratio}; $\kappa_t^e(v)$ and $\delta_t^l(u)$ represent the error functions associated with the partially observed and updating periods.

To estimate $\beta(u,v)$, we link two sets of principal component scores. Let $\bm{\widehat{\theta}}=[\bm{\widehat{\theta}}_{1},\bm{\widehat{\theta}}_{2},\dots,\bm{\widehat{\theta}}_{R}]$, where $\bm{\widehat{\theta}}_{r}=[\widehat{\theta}_{1,r},\widehat{\theta}_{2,r},\dots,\widehat{\theta}_{n,r}]^{\top}$ and $\bm{\widehat{\vartheta}}=[\bm{\widehat{\vartheta}}_{1},\bm{\widehat{\vartheta}}_{2},\dots,\bm{\widehat{\vartheta}}_{S}]$, where $\bm{\widehat{\vartheta}}_{s}=[\widehat{\vartheta}_{1,s},\widehat{\vartheta}_{2,s},\dots,\widehat{\vartheta}_{n,s}]^{\top}$. Via ordinary least squares, we study a linear relationship between $\bm{\widehat{\theta}}$ and $\bm{\widehat{\vartheta}}$, given by
\begin{equation}
\bm{\widehat{\vartheta}} = \bm{\widehat{\theta}}\times \bm{\rho},\label{eq:scores_relate}
\end{equation}
where $\bm{\rho}$ can be estimated by
\begin{equation}
\widehat{\bm{\rho}} = \big(\bm{\widehat{\theta}}^{\top}\bm{\widehat{\theta}}\big)^{-1}\bm{\widehat{\theta}}^{\top}\bm{\widehat{\vartheta}}.\label{eq:rho_estimate}
\end{equation}

The one-step-ahead point forecast of $\X_{n+1}^l(u)$ can be obtained by
\begin{equation}
\widehat{\X}_{n+1}^l(u) = \overline{\X}^l(u)+\sum^{S}_{s=1}\widehat{\vartheta}_{n+1,s}\widehat{\phi}_s^l(u).\label{eq:FLR}
\end{equation}
Given~\eqref{eq:scores_relate},~\eqref{eq:FLR} can be approximated by
\begin{equation}
\widehat{\X}_{n+1}^l(u)\approx \overline{\X}^l(u) + \widehat{\bm{\theta}}_{n+1}\times \widehat{\bm{\rho}}\times \bm{\widehat{\phi}}^l(u),\label{eq:FLR_2}
\end{equation}
where $\bm{\widehat{\phi}}^l(u) = \left[\widehat{\phi}_1^l(u), \widehat{\phi}_2^l(u), \dots,\widehat{\phi}_S^l(u)\right]$. 

The one-step-ahead interval forecast of $\X_{n+1}^l(u)$ can be obtained by
\[
\widehat{\X}_{n+1}^{l,*}(u) = \overline{\X}^l(u) + \int_{v\in \mathcal{I}_e}\left[\X_{n+1}^e(v) - \overline{\X}^e(v)\right]\widehat{\beta}^*(u, v)dv+e_{n+1}^{l,*}(u),
\]
where $\widehat{\beta}^*(u,v)$ represents the bootstrapped regression coefficient estimates, and $e_{n+1}^{l,*}(u)$ represents the bootstrapped residuals corresponding to the updating period, see~\eqref{eq:FPCA_boot}. Via the sieve bootstrap, we obtain bootstrap curves. By projecting these curves onto the estimated functional principal components, we obtain bootstrap principal component scores. Via~\eqref{eq:rho_estimate}, we obtain bootstrap $\bm{\rho}^*$, from which we form the bootstrap forecasts for the updating period via~\eqref{eq:FLR_2}.

The $(1-\alpha)$ prediction intervals for the updated forecasts are defined as $\alpha/2$ and $(1-\alpha/2)$ empirical quantiles of $\big\{\widehat{\X}_{n+1}^{l,1}(u), \widehat{\X}_{n+1}^{l,2}(u), \dots,\widehat{\X}_{n+1}^{l,B}(u)\big\}$.

\section{Evaluations of point and interval forecast accuracy}\label{sec:5}

In the forecasting literature, two schemes exist to evaluate forecast accuracy: the expanding and rolling windows. While a rolling window is a fixed size, expanding window has a fixed starting point and incorporates the most recent data as it becomes available. We consider the expanding-window scheme. The initial training samples are curves from days 1 to 200, and we compute a one-day-ahead forecast using the time-series method for day 201. Through an expanding window approach, we increase the training samples from days 1 to 201, and we then compute the one-day-ahead forecast for day 202. We iterate this procedure until the training samples cover the entire 250 days. 

\subsection{Mean squared forecast error}\label{sec:5.1}

We compute the out-of-sample point forecasts and evaluate their accuracy by mean squared forecast error (MSFE). The MSFE measures the closeness of the forecasts compared to the actual values of the variable being forecast.
\[
\text{MSFE}(u_i) = \frac{1}{n_{\text{test}}}\sum^{n_{\text{test}}}_{\iota=1}\left[\X_{\iota}(u_i) - \widehat{\X}_{\iota}(u_i)\right]^2 \qquad
\text{MSFE} = \frac{1}{\tau-1}\sum^{\tau}_{i=2}\text{MSFE}(u_i)
\]
where $\X_{\iota}(u_i)$ represents the actual holdout samples for the $i$\textsuperscript{th} intraday period on the $\iota$\textsuperscript{th} day, $\widehat{\X}_{\iota}(u_i)$ represents the one-step-ahead point forecasts for the holdout samples, and $n_{\text{test}}=50$ represents the number of curves in the out-of-sample forecasting period. 

\subsection{Empirical coverage probability and interval score}\label{sec:5.2}

To evaluate interval forecast accuracy, we compute empirical coverage probabilities (ECPs) for pointwise prediction intervals and uniform prediction bands. The ECPs are defined by
\begin{align*}
\text{ECP}_{\text{pointwise}} &= 1 - \frac{1}{n_{\text{test}}\times (\tau-1)}\sum^{n_{\text{test}}}_{\iota=1}\sum_{i=2}^{\tau} \left[\mathds{1}\{\X_{\iota}(u_i)<\widehat{\X}_{\iota}^{\text{lb}}(u_i)\} + \mathds{1}\{\X_{\iota}(u_i) > \widehat{\X}_{\iota}^{\text{ub}}(u_i)\}\right] \\
\text{ECP}_{\text{uniform}} &= 1-\frac{1}{n_{\text{test}}}\sum^{n_{\text{test}}}_{\iota=1}\left[\mathds{1}\{\X_{\iota}(u)<\widehat{\X}_{\iota}^{\text{lb}}(u)\}+\mathds{1}\{\X_{\iota}(u) > \widehat{\X}_{\iota}^{\text{ub}}(u)\}\right].
\end{align*}

The uniform prediction bands are wider than the pointwise prediction intervals. As a result, the uniform empirical coverage probability counts the number of days in the forecasting period, where the constructed prediction bands cover all future realizations.

For pointwise prediction intervals, we also utilize the interval score of \cite{GR07} and \cite{GK14}. For each year in the forecasting period, the one-step-ahead prediction intervals were computed at the $(1-\alpha)$ nominal coverage probability. Let $\widehat{\X}_{\iota}^{\text{lb}}(u_i)$ and $\widehat{\X}_{\iota}^{\text{ub}}(u_i)$ be the lower and upper bounds of the pointwise prediction intervals, respectively. A scoring rule for the prediction interval at time point $u_i$ is
\begin{align*}
S_{\alpha}\left[\widehat{\X}_{\iota}^{\text{lb}}(u_i), \widehat{\X}_{\iota}^{\text{ub}}(u_i), \X_{\iota}(u_i)\right]=\left[\widehat{\X}_{\iota}^{\text{ub}}(u_i) - \widehat{\X}_{\iota}^{\text{lb}}(u_i)\right] &+ \frac{2}{\alpha}\left[\widehat{\X}^{\text{lb}}_{\iota}(u_i) - \X_{\iota}(u_i)\right]\mathds{1}\left\{\X_{\iota}(u_i)<\widehat{\X}_{\iota}^{\text{lb}}(u_i)\right\} \\
& + \frac{2}{\alpha}\left[\X_{\iota}(u_i) - \widehat{\X}_{\iota}^{\text{ub}}(u_i)\right]\mathds{1}\left\{\X_{\iota}(u_i)>\widehat{\X}_{\iota}^{\text{ub}}(u_i)\right\}.
\end{align*}

Averaged over different days in the forecasting period, the mean interval score for each intraday period $i$ and its averaged mean interval score are defined by
\begin{align*}
\overline{S}_{\alpha}(u_i) &= \frac{1}{n_{\text{test}}}\sum^{n_{\text{test}}}_{\iota=1}S_{\alpha}\left[\widehat{\X}_{\iota}^{\text{lb}}(u_i), \widehat{\X}_{\iota}^{\text{ub}}(u_i), \X_{\iota}(u_i)\right], \\
\overline{S}_{\alpha} &= \frac{1}{\tau-1}\sum^{\tau}_{i=2}\overline{S}_{\alpha}(u_i).
\end{align*}

\section{Results}\label{sec:6}

\begin{figure}[!htb]
\centering
\includegraphics[width=16.9cm]{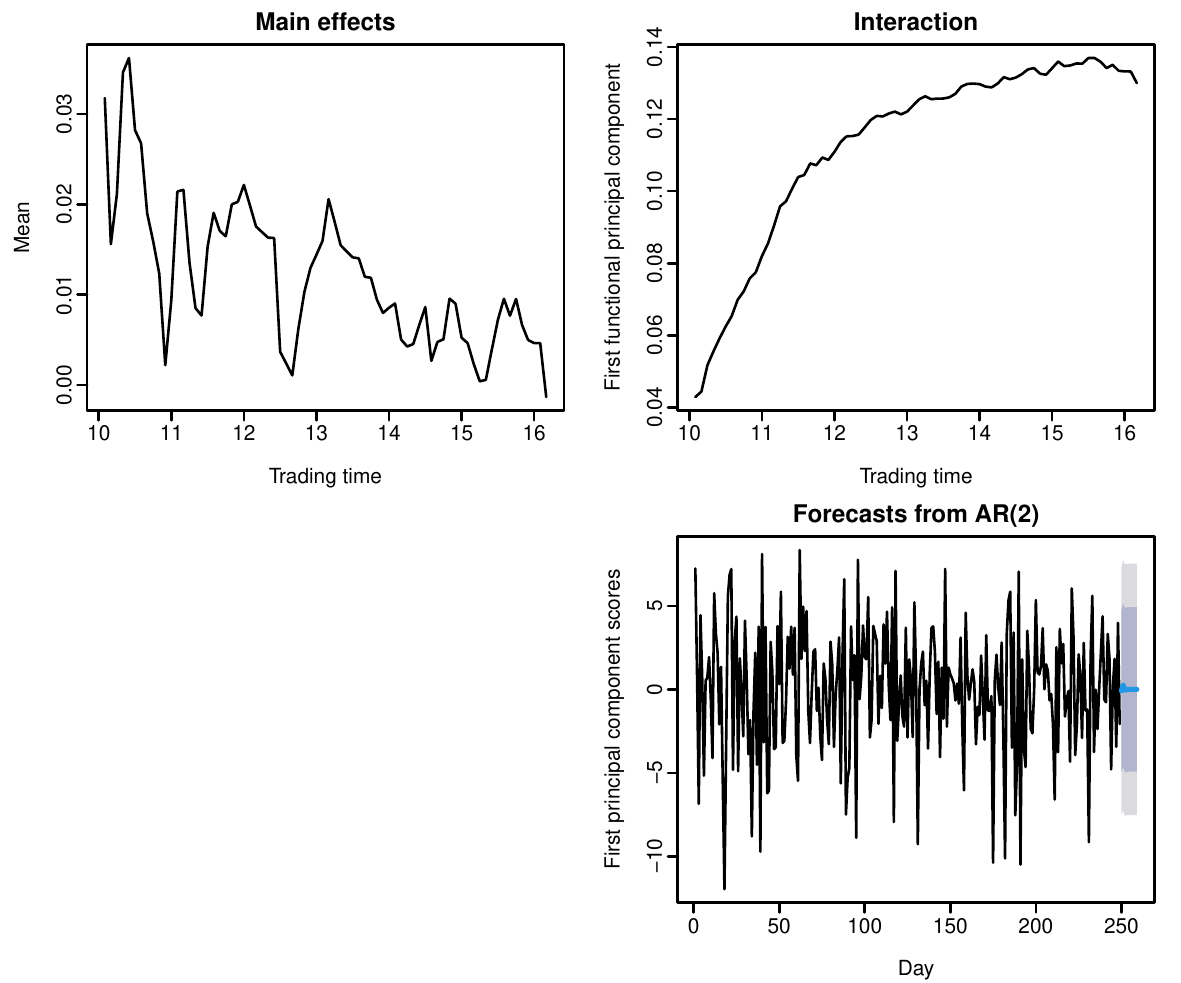}
\caption{Estimated mean function, first set of functional principal component, and its associated principal component scores for the CIDRs of the XAO from January 4 to December 22, 2021. For the ten-day-ahead forecasts, the 80\% and 95\% prediction intervals of the estimated principal component scores are shown by the dark and light gray regions. For the univariate time series of the principal component scores, the AR(2) is selected as the optimal model based on the Akaike information criterion.}\label{fig:3}
\end{figure}

The functional time series forecasting method in Section~\ref{sec:3} decomposes a time series of functions, i.e., CIDRs from January 4 to December 22, 2021, into a set of functional principal components and their associated principal component scores. The temporal dependence of the functional time series is inherited by the correlation among the principal component scores. From~\eqref{eq:eigenvalue_ratio}, the retained number of components is one and the selected order of the VAR model is $p=4$ in~\eqref{eq:AICC}. We display and attempt to interpret the first functional principal component in the top panel of Figure~\ref{fig:3}. The mean function illustrates average changes in CIDRs with larger fluctuations at or before 14:00 and relatively smaller fluctuations after 14:00. The first functional principal component shows a general increasing trend in CIDRs. As a demonstration, the ten-days-ahead point forecasts of the principal component scores show an increasing trend for a short-time interval before stabilizing. This indicates the short-term predictive ability of the CIDRs. 

\begin{figure}[!htb]
\centering
\includegraphics[width=12cm]{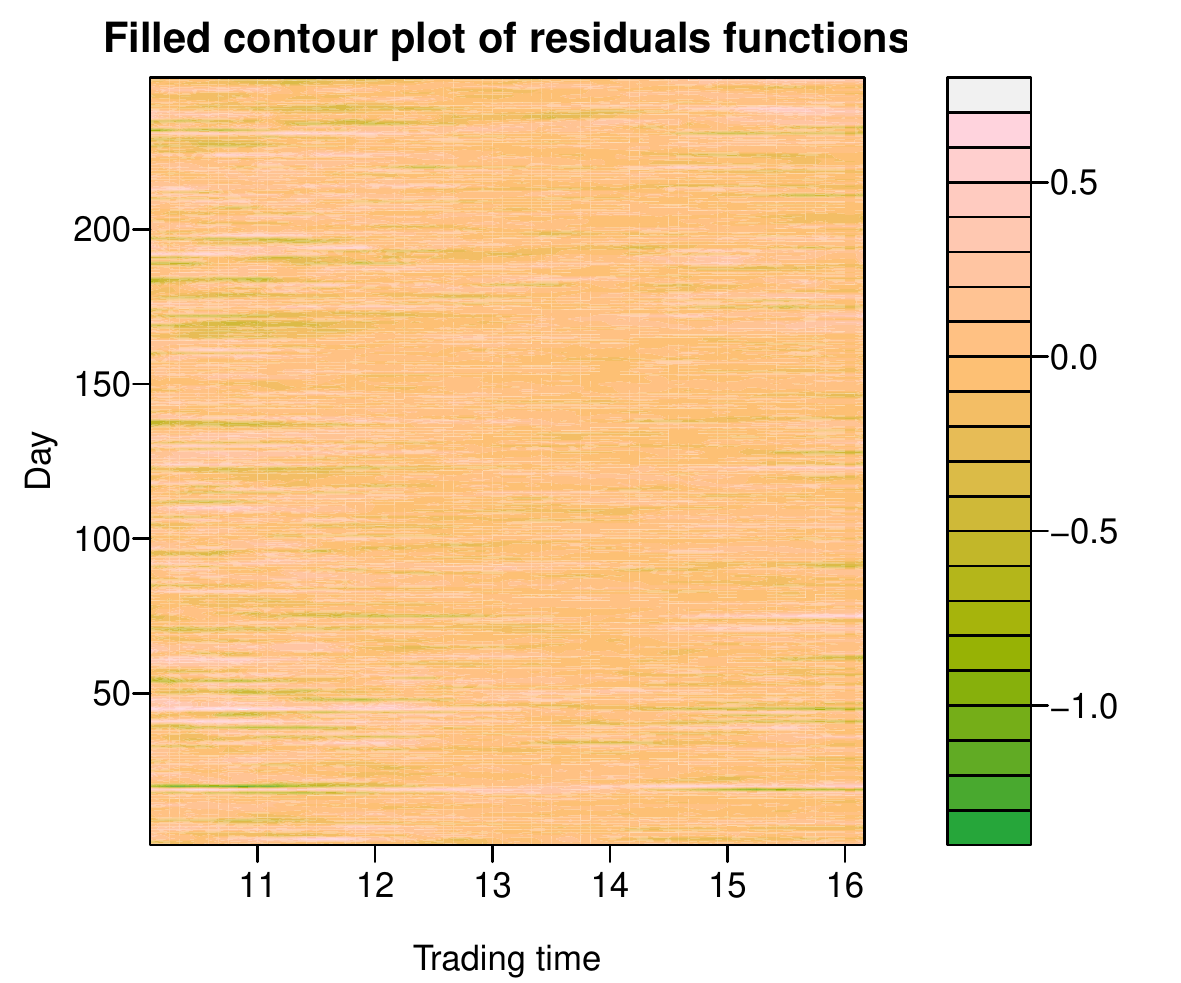}
\caption{Filled contour plot of the residual functions obtained from the fitted functional time series model. Using a terrain color palette, negative residuals are shown in green, positive residuals in gray and pink, and residuals around zero are shown in orange.}\label{fig:4}
\end{figure}

To assess overall goodness-of-fit, the residual functions of the fitted functional time series are displayed in Figure~\ref{fig:4} via a filled contour plot. As no clear pattern is exhibited in the contour plot, the fitted model captures the temporal structure well. Further, we implement a white noise test of \cite{GK07} to check if there is any temporal correlation in the residual functions. From the $p$-value of 0.4453, we conclude that the residuals functions are white noise.

\begin{figure}[!htb]
\centering
\includegraphics[width=11.5cm]{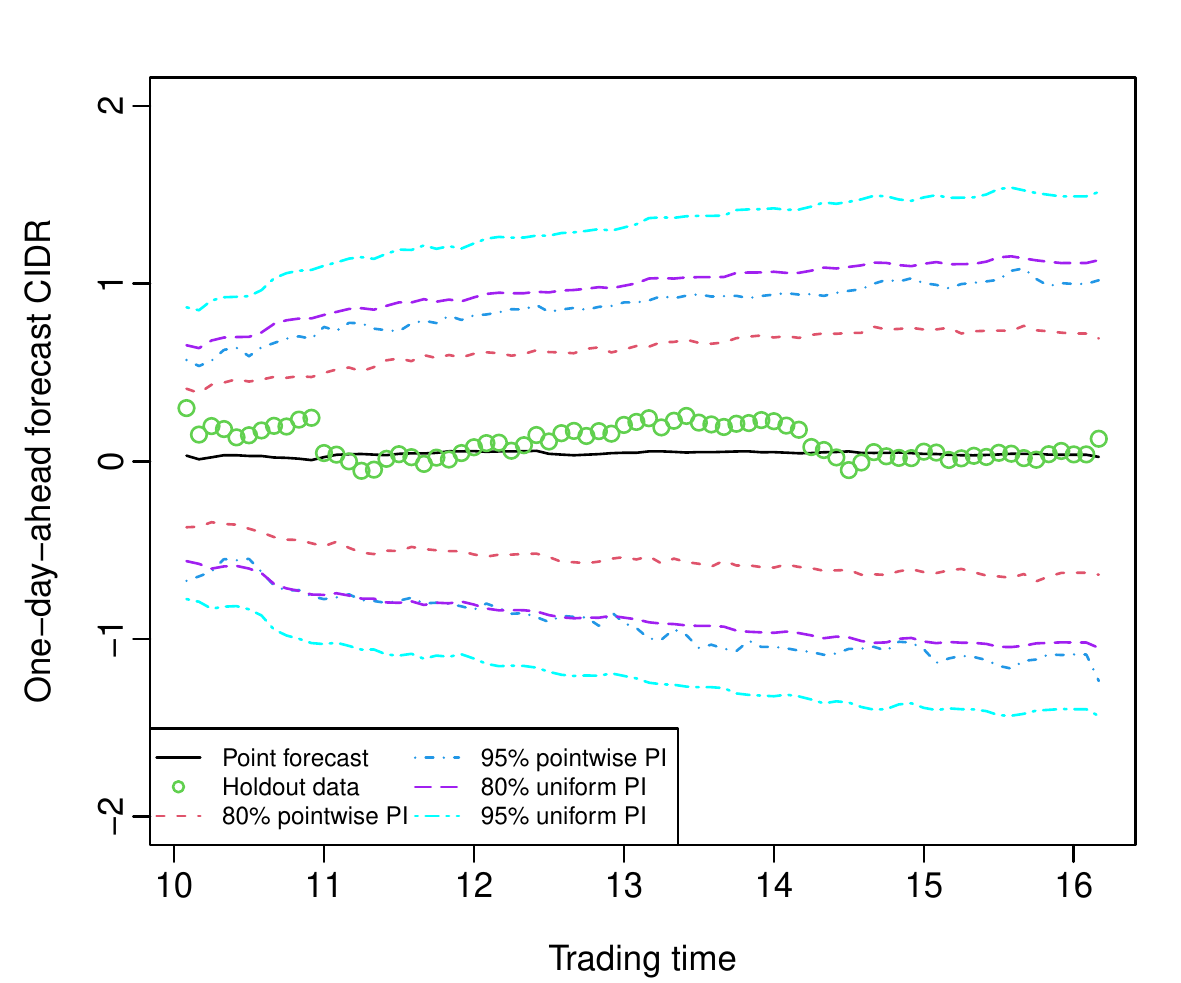}
\caption{One-step-ahead point forecasts of the CIDRs of the XAO on December 23, 2021, the 80\% and 95\% pointwise prediction intervals and uniform prediction bands constructed via the sieve bootstrap.}\label{fig:5}
\end{figure}

As an illustration, we model CIDRs from January 4 to December 22, 2021, and forecast the CIDR for December 23, 2021. We multiply the one-day-ahead forecast principal component scores by the estimated functional principal components and add the estimated mean function. Using the sieve bootstrap method described in Section~\ref{sec:3.2}, the 80\% and 95\% pointwise prediction intervals and uniform prediction bands, shown in Figure~\ref{fig:5}, are constructed from these bootstrapped forecasts.

Averaged over 50 days in the forecasting period and 73 different intraday updating periods, we present the averaged MSFE, ECP$_{\text{pointwise}, 1-\alpha}$, ECP$_{\text{uniform}, 1-\alpha}$ and $\overline{S}_{\alpha}$ obtained from the TS method. In Table~\ref{tab:1}, the sieve bootstrap method produces the empirical coverage probabilities that are no less than the nominal coverage probabilities. 

\begin{table}[!htb]
\centering
\caption{Averaged forecast errors of the TS method over 50 days in the forecasting period.}\label{tab:1}
\tabcolsep 0.14in
\begin{tabular}{@{}llcccccc@{}}
\toprule
& & \multicolumn{2}{c}{ECP$_{\text{pointwise},1-\alpha}$} & \multicolumn{2}{c}{ECP$_{\text{uniform},1-\alpha}$} & \multicolumn{2}{c}{$\overline{S}_{\alpha}$} \\ 
\cmidrule{3-8}
Method & MSFE & $\alpha = 0.2$ & $\alpha = 0.05$ & $\alpha = 0.2$ & $\alpha = 0.05$ & $\alpha = 0.2$ & $\alpha = 0.05$ \\
\midrule
TS & 0.1474 & 0.89 & 0.98 & 0.88 & 0.96 & 1.43 & 2.08 \\
\bottomrule
\end{tabular}
\end{table}

\subsection{Updating point forecasts}\label{sec:6.1}

\begin{figure}[!htb]
\centering
\subfloat[Selected optimal $\lambda$ values]
{\includegraphics[width=8.3cm]{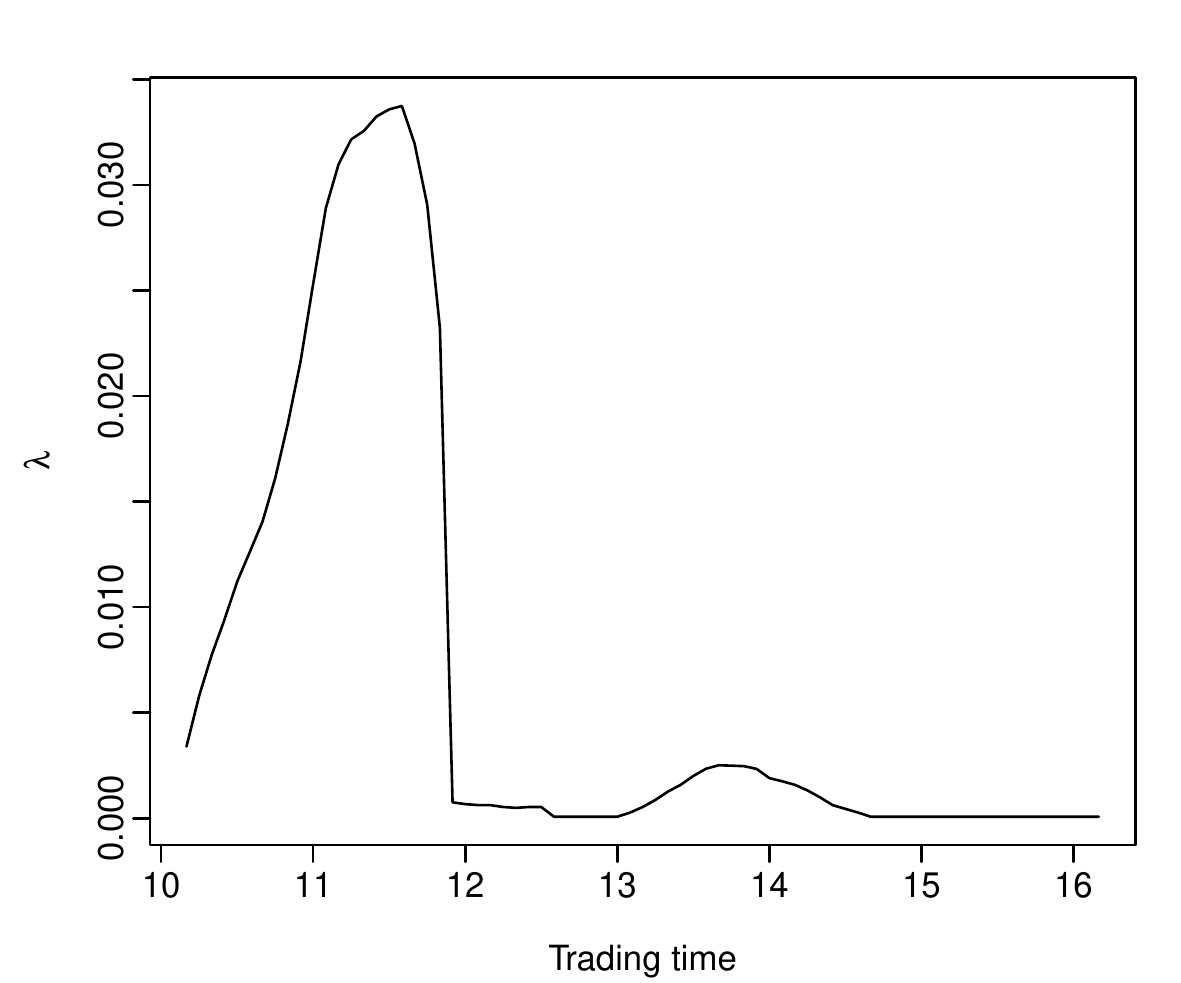}\label{fig:6a}}
\subfloat[Out-of-sample MSFE]
{\includegraphics[width=8.3cm]{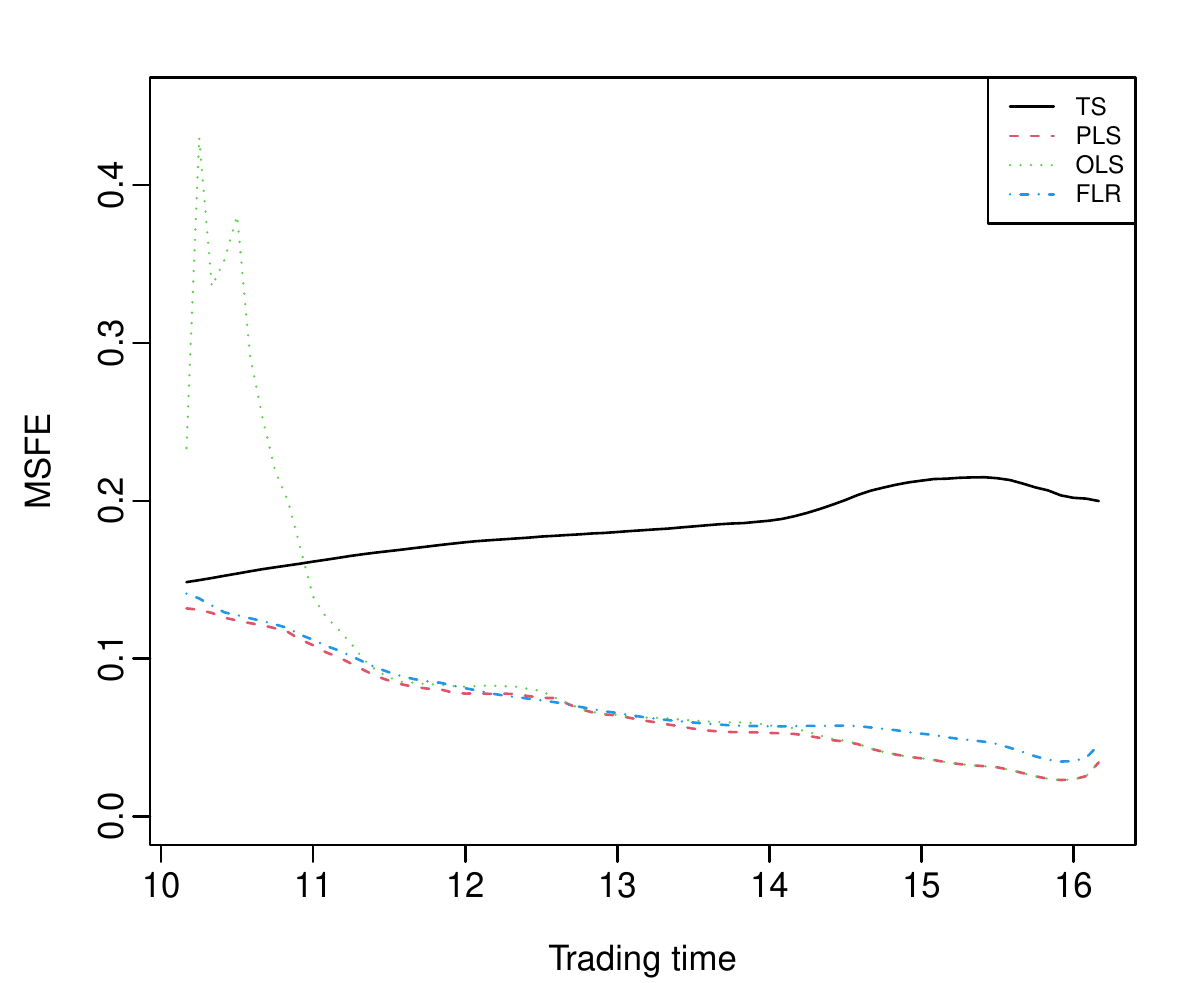}\label{fig:6b}}
\caption{A comparison of point forecast accuracy, as measured by averaged $\text{MSFE}_{i}$ for $i = 3, 4, \dots,\tau$ over 50 days in the forecasting period, among the TS, PLS, OLS, and FLR methods. We observe at least one data point in the dynamic updating methods before updating forecasts.}\label{fig:6}
\end{figure}

When we have partially observed data in the most recent curve, we can dynamically update our point forecasts for the remaining period to improve forecast accuracy. Two dynamic updating methods were presented in Section~\ref{sec:4}, and their point forecast accuracies are evaluated and compared in Figure~\ref{fig:6b}. The forecast errors generally decrease throughout the day for the dynamic updating methods, as we have more partially observed data in the most recent curve. Averaged over 50 days in the forecasting period, the PLS method has the best point forecast accuracy, measured by MSFE, over all evaluation intraday time points. When $\lambda=0$ in~\eqref{eq:PLS_2}, the PLS method reduces to the OLS method. The PLS forecasts produce smaller errors than the OLS forecasts in the early trading time, and then their forecasts and forecast errors are similar in the afternoon. This phenomenon is reflected by the estimated optimal values of $\lambda$. As shown in Figure~\ref{fig:6a}, the $\lambda$ values are selected by minimizing the MSFE within the validation set. 

\begin{figure}[!htb]
\centering
\includegraphics[width=9.4cm]{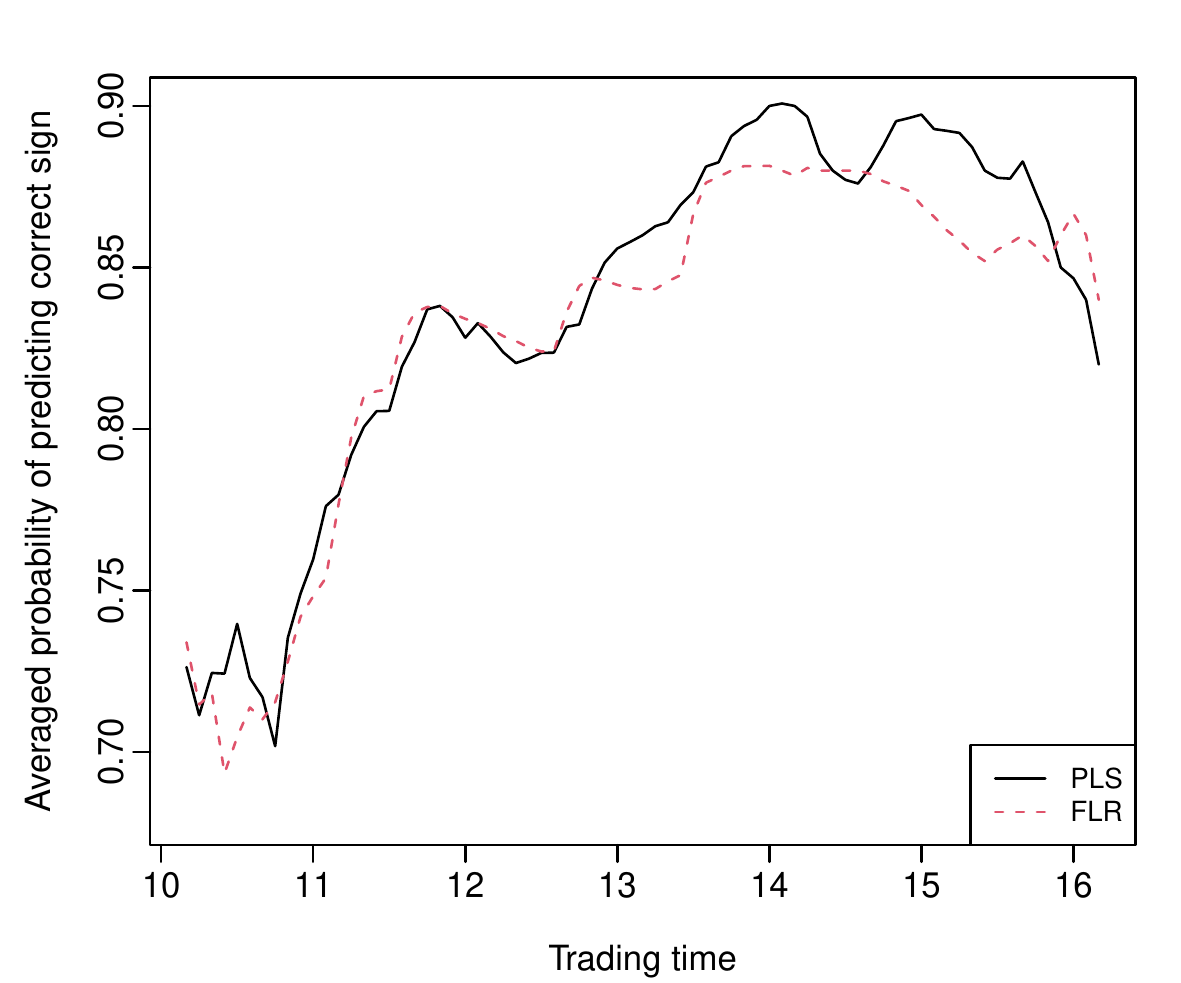}
\caption{Averaged over 50 days in the forecasting period, the probabilities of predicting the correct sign are computed for the PLS and FLR methods at different updating periods.}\label{fig:7}
\end{figure}

Future values of the CIDRs are hard to predict accurately, but the signs of future values are relatively easier \citep[see also][]{HLM92}. Averaged over 50 days in the forecasting period, we count the number of times where forecasts and holdout values share the same sign, i.e., negative or positive. Using the PLS method, we compute the probabilities of making correct sign predictions over different updating periods in Figure~\ref{fig:7}. 

The probability of correct sign prediction increases from 0.7263 at the beginning of the updating period (i.e., 10:15-16:10) to 0.8200 at the end of the intraday trading time (i.e., 16:10), a peak at 0.9008 occurring at 14:10-16:10. The peak occurs when the signs of the remaining observations are most accurately predicted on average. We also present the correct sign prediction probability using the FLR method, which generally performs inferior to the PLS method. A plausible explanation for the inferior performance of the FLR method is that it does not use the TS regression coefficient as a priori. 

As a trading strategy, a day trader can enter the market at around 14:10 since we can predict the direction more accurately. For example, if the predictions indicate the XAO will likely go up, the day trader may choose to follow the momentum and sell it at or close to the closing trading time of the day. The daily profit can be made so long the difference between purchasing and selling prices covers transaction costs. 

\subsection{Updating interval forecasts}\label{sec:6.2}

\begin{figure}[!htb]
\centering
\subfloat[80\% nominal coverage probability]
{\includegraphics[width=8cm]{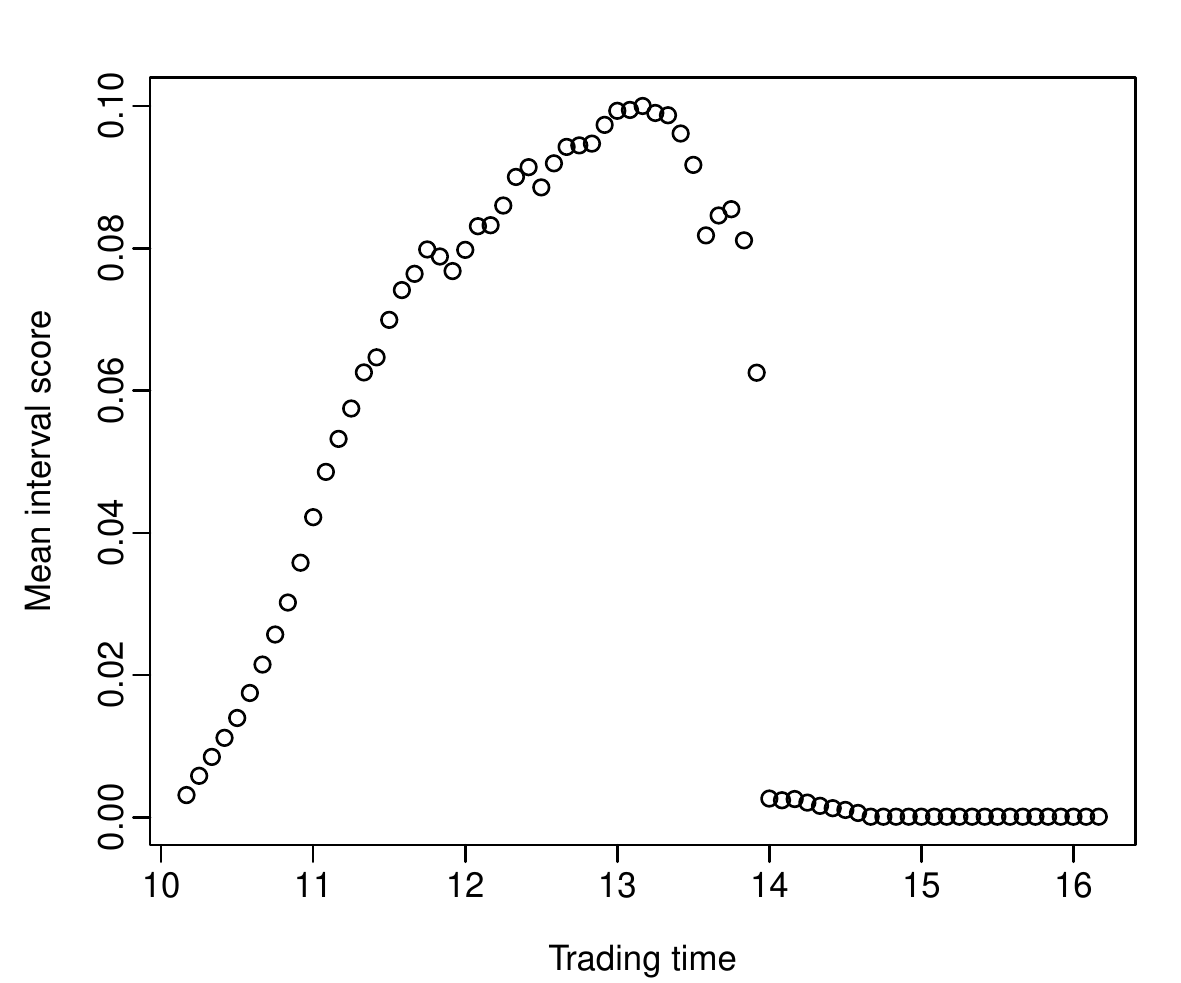}}
\quad
\subfloat[95\% nominal coverage probability]
{\includegraphics[width=8cm]{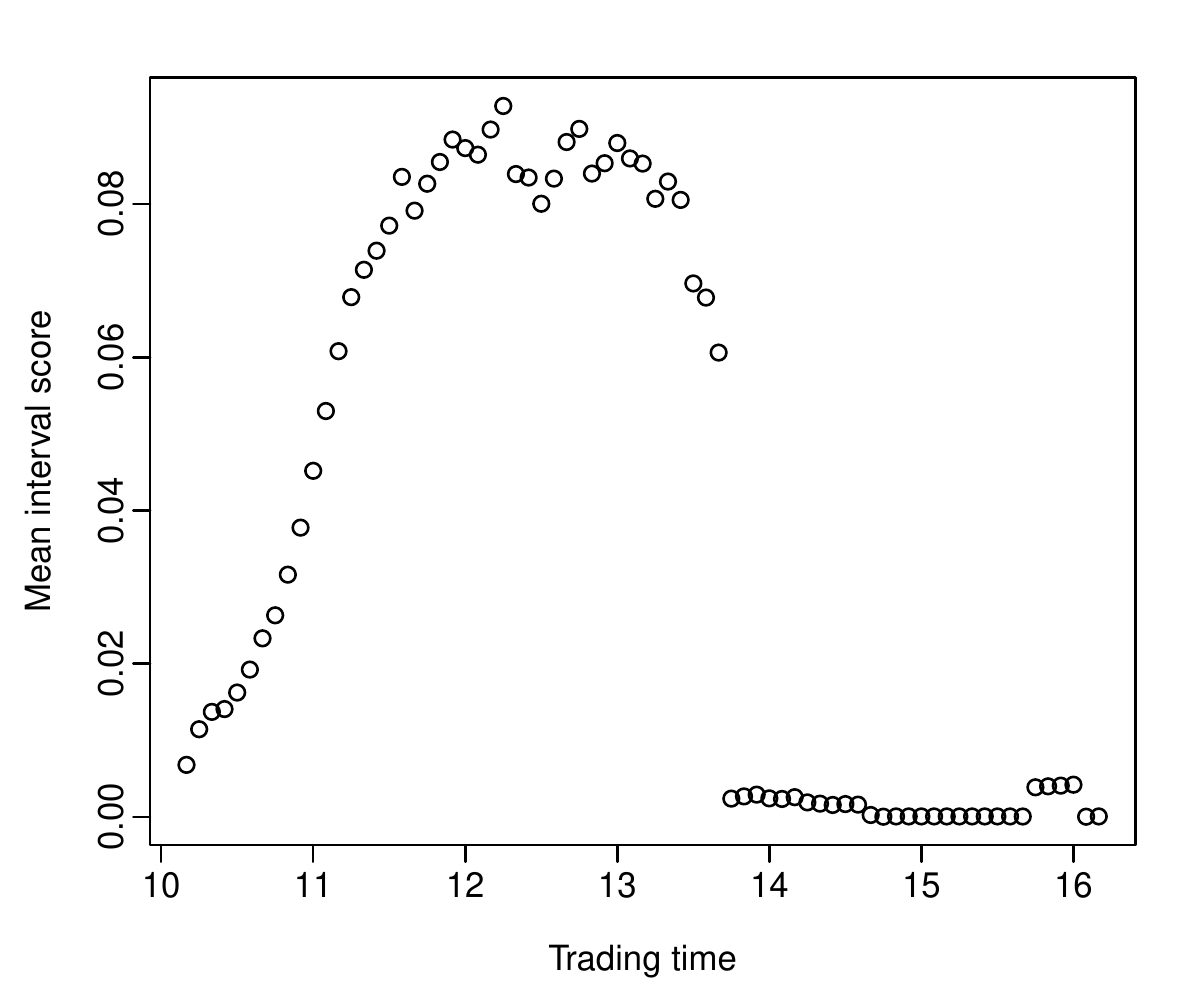}}
\caption{Estimated optimal values of $\lambda$ for different intraday updating periods by minimizing the mean interval scores within the validation set.}\label{fig:8}
\end{figure}

We can also update our pointwise interval forecasts for the remaining period to achieve improved accuracy when we have partially observed data in the most recent curve. We estimate the optimal values of $\lambda$ for different updating periods by minimizing the averaged mean interval score within the validation set for the PLS method. The optimal values are presented in Figure~\ref{fig:8} for the 80\% and 95\% nominal coverage probabilities. At the 80\% nominal coverage probability, the values of $\lambda$ increase from the beginning of the trading time to around 13:15, then decrease sharply from 14:00 to 16:10. At the 95\% nominal coverage probability, the values of $\lambda$ increase from the beginning of the trading time to 12:15, stabilize until 13:30 and decrease sharply from 13:40 to 16:10.

\begin{figure}[!htb] 
\centering
\includegraphics[width=11cm]{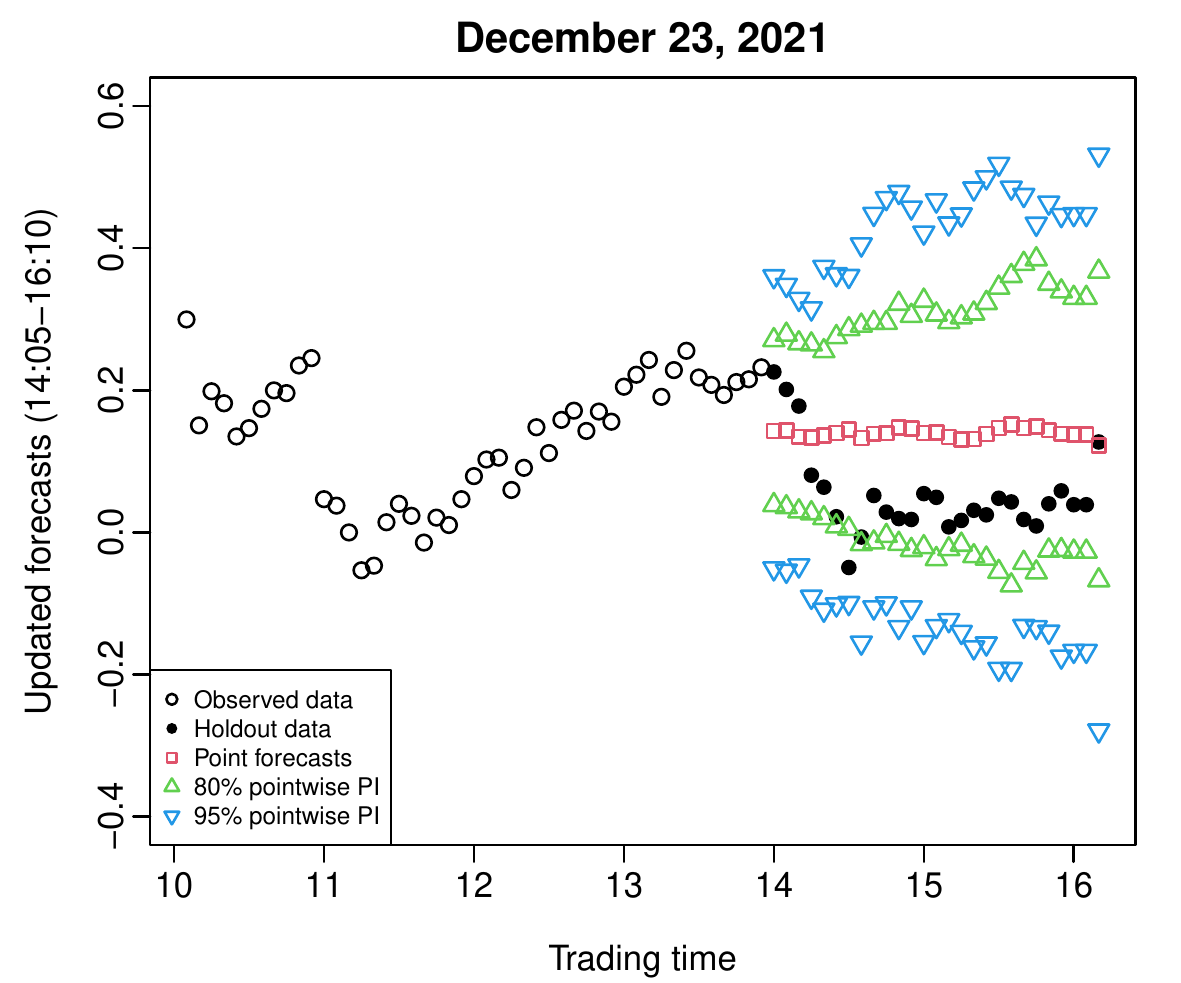}
\caption{As we sequentially observe new information from the beginning of the trading time to 14:00, we apply the PLS method with optimally tuned parameters to update point and interval forecasts of the CIDRs of the XAO on December 23, 2021.}\label{fig:9}
\end{figure}

Suppose the CIDRs of the XAO have been observed from 10:05 to 14:00 on December 23, 2021, we apply the PLS and FLR methods to update the interval forecasts for the remaining period of that day. Figure~\ref{fig:9} presents the 80\% and 95\% pointwise prediction intervals when the dynamic updating methods are used for the remaining period. 

\begin{figure}[!htb]
\centering
\includegraphics[width=7.95cm]{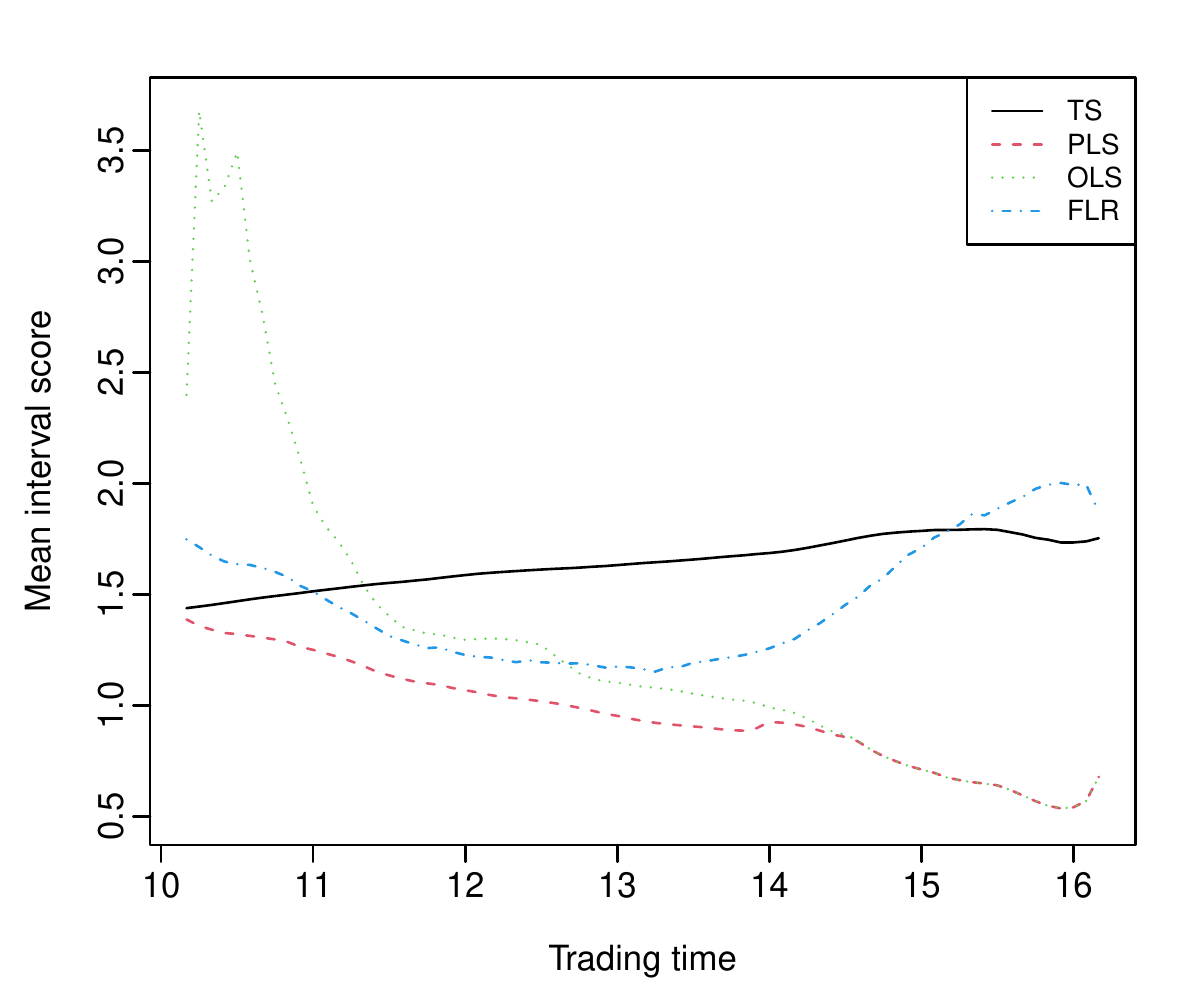}
\quad
\includegraphics[width=7.95cm]{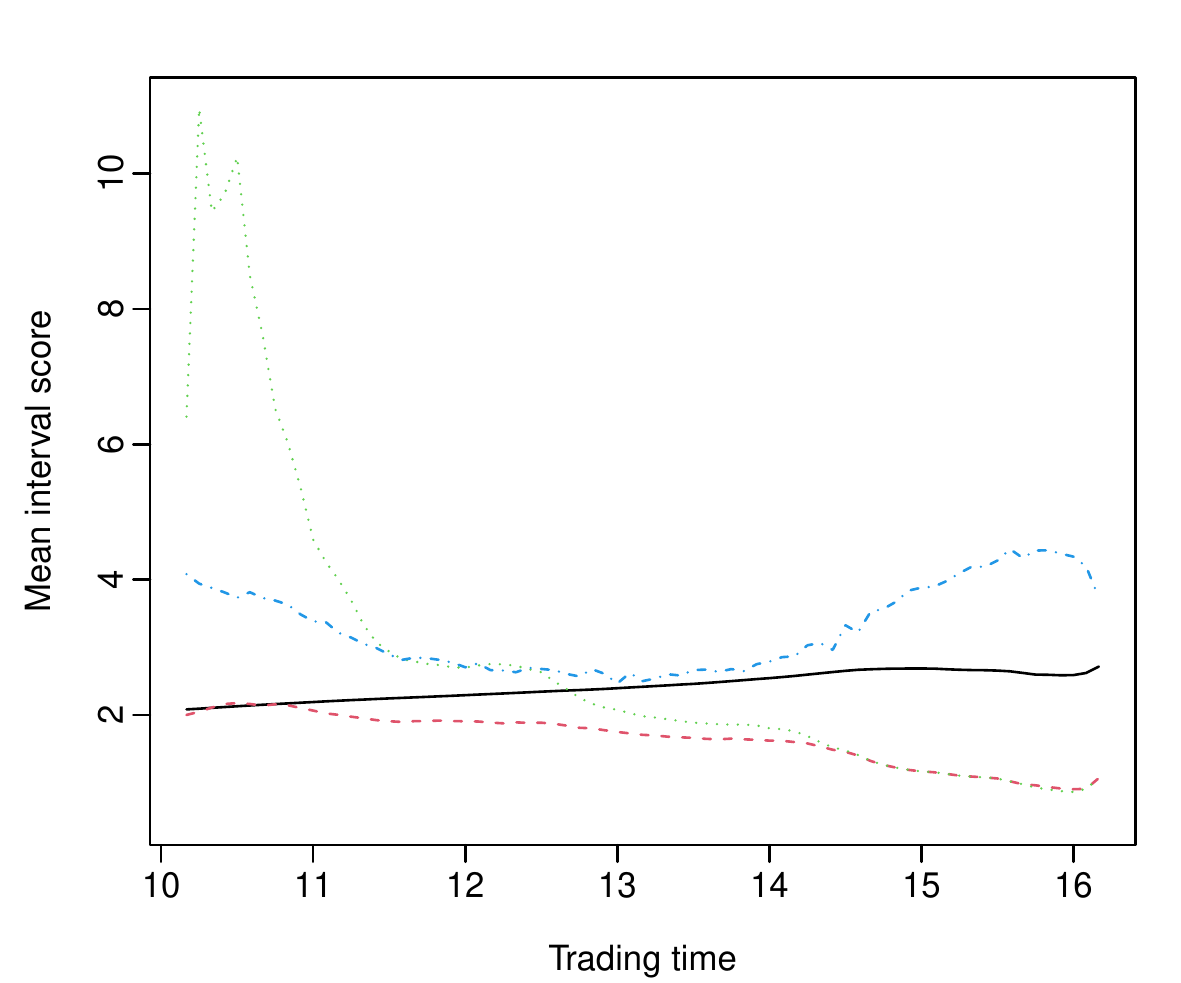}
\caption{A comparison of interval forecast accuracy, as measured by averaged mean interval score for different updating periods over 50 days in the forecasting period, among the TS, PLS, OLS, and FLR methods.}\label{fig:10}
\end{figure}

Between the PLS and FLR methods, as shown in Figure~\ref{fig:10}, the PLS method produces the smaller mean interval score, averaged over 50 days in the forecasting period. Thus, it provides the most accurate evaluation of forecast uncertainty. 

Averaged over 50 days in the forecasting period and 73 different intraday updating periods, we present the averaged MSFE, ECP$_{\text{pointwise},1-\alpha}$, ECP$_{\text{uniform},1-\alpha}$ and $\overline{S}_{\alpha}$ obtained from the TS, PLS, OLS and FLR methods. In Table~\ref{tab:2}, the dynamic updating methods, such as the PLS and FLR, improve the point forecast accuracy as measured by the MSFE. In terms of the interval forecast accuracy, the ECPs of the PLS method are smaller than the nominal ones, which may be due to the same estimated values of $\lambda$ are used in each bootstrap sample. The lack of uncertainty associated with $\lambda$ stems from its computational cost and trade-off between accuracy and practicality. Nevertheless, the PLS method produces the smallest MSFE and $\overline{S}_{\alpha}$, and thus it is our recommended method.
\begin{table}[!htb]
\centering
\caption{By averaging 73 different intraday updating periods, we compute the averaged forecast errors of the TS, PLS, OLS, and FLR methods over 50 days in the forecasting period.}\label{tab:2}
\tabcolsep 0.32in
\begin{tabular}{@{}lccccc@{}}
\toprule
& & \multicolumn{2}{c}{ECP$_{\text{pointwise},1-\alpha}$} & \multicolumn{2}{c}{$\overline{S}_{\alpha}$} \\ 
\cmidrule{3-6}
Method & MSFE & $\alpha = 0.2$ & $\alpha = 0.05$ & $\alpha = 0.2$ & $\alpha = 0.05$ \\
\midrule
TS & 0.1838 & 0.8817 & 0.9688 & 1.6429 & 2.4277 \\
PLS & 0.0672 & 0.7275 & 0.8851 & 0.9564 & 1.6440 \\
FLR & 0.0735 & 0.5001 & 0.7266 & 1.4661 & 3.2729 \\
\bottomrule
\end{tabular}
\end{table}

\section{Conclusion}\label{sec:7}

Our forecasting method treats the observations as realizations of a functional time series, where the temporal dependence among curves can be modeled and forecasted via functional principal component analysis. The temporal dependence in the functional time series is manifested in the estimated principal component scores. By explaining at least 85\% of the total variation in the data, we select the retained number of components. When the number of components is one, a multivariate time series technique, such as VAR($p$) model, is reduced to a univariate time series technique, such as AR($p$) model. Conditioning on the estimated functional principal components and historical curves, a one-step-ahead point forecast can be obtained. 

When partial data in the most recent curve are sequentially observed, two dynamic updating methods, the PLS and FLR methods, are implemented to update forecasts. The FLR method decomposes two blocks of functions via functional principal component analyses. The first block is associated with the data in the partially observed period of the most recent curve; the second block is associated with the data in the updating period. By linking these estimated principal component scores via linear regression, a regression coefficient estimate for these scores is obtained. Conditioning on the estimated functional principal components and observed data in the updating period, updated forecasts can be obtained based on the estimated regression coefficient of the scores and the principal component scores in the partially observed data of the most recent curve. The PLS method is a shrinkage estimator: When the newly arrived data occur in the morning of the trading time, it shrinks its regression coefficient towards time-series regression coefficient with a comparably larger tuning parameter $\lambda$. When the newly arrived data cover most of the day, it shrinks its regression coefficient towards the OLS regression coefficient. By selecting the optimal values of $\lambda$, the PLS method achieves superior forecast accuracy among all methods considered for producing point and interval forecasts, measured by averaged MSFE and interval score, respectively. 

As a means of measuring forecast uncertainty, a sieve bootstrap method was used to construct pointwise prediction intervals and uniform prediction bands for the TS method. The sieve bootstrap method takes into account model misspecification error and thus achieves excellent calibration, where the empirical coverage probability is no less than the nominal coverage probability. The sieve bootstrap plays an important role in constructing pointwise prediction intervals for the updated forecasts using the PLS and FLR methods.

There are at least three ways in which the present paper can be extended: 
\begin{inparaenum}
\item[1)] While the methods apply to five-minute intervals, future research could consider optimal sampling frequency, possibly even shorter-time intervals. 
\item[2)] The presence of outliers can affect the estimation of the covariance structure. Since the covariance relies on the sample mean, it affects the estimation accuracy of the functional principal components. Consequently, it may affect the accuracy of forecasts, and one could consider robust functional principal component analysis \citep[see, e.g.,][]{BBT+11}. 
\item[3)] With a set of validation samples, the PLS tuning parameters can be adaptively chosen without re-computing them.
\end{inparaenum}

\section*{Acknowledgments}

The authors thank a reviewer for insightful comments and suggestions. This study was partially supported by an Australian Research Council Discovery Project (ARC DP230102250). The author acknowledges seminar participants at the Department of Applied Finance, Macquarie University, for their insightful comments and suggestions.


\end{document}